
%
%

\input phyzzx

\overfullrule=0pt


\def\pt{\partial}
\def\S{\Sigma}

\def\sym #1/#2/#3/#4/#5/#6/{
     \left|  \matrix{#1 & #2 & #3 \cr #4 & #5 & #6 \cr}  \right|}


\FIG\figglu{Two lattices $M_1$ and $M_2$ with boundaries
are glued at sub-boundaries $b$.}

\FIG\figslfglu{A lattice $M_2$ is obtained by identifying
sub-boundaries $b$ of $M_1$.}

\FIG\figtwosim{Two lattices with $\pt {\cal T}^D$ as their
simplically decomposed boundaries
are glued at one of their ${\cal T}^{D-1}$s of each boundary.
This figure is for $D=3$.}

\FIG\figflip{The 2D lattice QCD is invariant under the flip move,
and hence is an area-dependent TLFT.}

\FIG\figdifman{If the $\delta Z$s come from the $D$-simplices
shown in the figure, to estimate the contributions to the
partition function,
one has to use correlation functions of the TLFT on
manifolds different from $M$.}

\FIG\figcluws{If the set $Q$ is separated from
the boundaries and the compliment $\bar Q$ near the boundaries,
then $W_n^\S$ is zero.}

\FIG\figtwobou{If the two boundaries are separated,
the lower order terms of $W_n^{\S_{12}}$ factorize.}

\FIG\figmove{The (1,4) and (2,3) moves of the simplicial
decomposition and their inverses.
In three dimensions, all the simplicial decompositions
of a manifold can be generated by these moves.}

\FIG\figgeo{The geometrical condition of $t\times B_g$ $(t\in S^1)$.}

\FIG\figopegam{$\Gamma_{kl}^{ij}$ is a physical state on the torus
winding through the two neighboring holes $k$ and $l$.}


\def\PRL{Phys.~Rev.~Lett.}
\def\CMP{Commun.~Math.~Phys.}
\def\PL{Phys.~Lett.}
\def\MPL{Mod.~Phys.~Lett}
\def\NP{Nucl.~Phys.}
\def\PMI{Publ.~Math.~IHES}
\def\JETP{Sov.~Phys.~J.E.T.P.}
\def\NC{Nuovo Cimento}
\def\TOP{Topology}
\def\AM{Ann.~Math.}
\def\RMP{Rev.~Math.~Phys.}
\def\PTP{Prog.~Theor.~Phys.}
\def\IJMP{Int.~J.~Mod.~Phys.}
\def\LMP{Lett.~Math.~Phys.}
\def\PR{Phys.~Rev.}
\def\CRASP{C.R.~Acad.~Sci.~Paris}
\def\JDG{J.~Diff.~Geom.}

\REF\WITONE{E.~Witten \journal \CMP &117 (88) 353.}

\REF\ATI{M.~Atiyah \journal \PMI &68 (89) 175.}

\REF\WITTWO{E.~Witten \journal \NP &B340 (90) 281.}

\REF\WITTHR{E.~Witten \journal \NP &B311 (88) 46.}

\REF\MOOSEI{G.~Moore and N.~Seiberg \journal \PL &B220 (89) 422.}

\REF\WITFOU{E.~Witten \journal \NP &B322 (89) 629;
E.~Witten \journal \NP &B330 (90) 285.}

\REF\WITFIV{E.~Witten \journal \CMP &121 (89) 351.}

\REF\WITSIX{E.~Witten \journal \CMP &117 (88) 353.}

\REF\WHEONE{J.F.~Wheater \journal \PL &B223 (89) 451.}

\REF\WHETWO{J.F.~Wheater \journal \PL &B264 (91) 161.}

\REF\JONONE{T.~J{\`o}nsson \journal \PL &B265 (91) 141.}

\REF\JONTWO{T.~J{\`o}nsson \journal \NP (Proc.~Suppl.)
&B 25A (92) 176.}

\REF\BACPET{C.~Bachas and P.M.S.~Petropoulos \journal \CMP &152 (93)
191.}

\REF\BORFIL{M.~Bordemann, T.~Filk and C.~Nowak, ``Topological actions
on 2-dimensional graphs and (graded) metrised algebras'', preprint
THEP 92/18.}

\REF\FUKKAW{M.~Fukuma, S.~Hosono and H.~Kawai, ``Lattice topological
field theory in two-dimensions'',
preprint CLNS~92/1173.}

\REF\DURJON{B.~Durhuus and T.~J{\`o}nsson, ``Classification and
construction of unitary topological field theories in two
dimensions'', preprint COPENHAGEN-MI-14-1993.}

\REF\MIGONE{A.~Migdal \journal \JETP &42 (75) 413.}

\REF\EGUYAN{T.~Eguchi and S.K.~Yang \journal \MPL &A5 (91) 1693.}

\REF\PONREG{G.~Ponzano and T.~Regge, in {\it Spectroscopic and Group
Theoretical Methods in Physics, }ed. F.~Bloch (North-Holland,
Amsterdam, 1968). }

\REF\REG{T.~Regge \journal \NC &19 (61) 558.}

\REF\TURVIR{V.G.~Turaev and O.Y.~Viro \journal \TOP &31 (92) 865.}

\REF\DIJWIT{R.~Dijkgraaf and E.~Witten \journal \CMP &129 (90) 393.}

\REF\DURJAC{B.~Durhuus, H.~Jacobsen and R.~Nest
\journal \NP
(Proc.~Suppl.) &25A (92) 109; B.~Durhuus, H.~Jacobsen
and R.~Nest\journal \RMP &5 (93) 1.}

\REF\OOGSAS{H.~Ooguri and N.~Sasakura \journal \MPL &A6 (91) 3591.}

\REF\BOU{D.V.~Boulatov \journal \MPL &A7 (92) 1629.}

\REF\ARCWIL{F.~Archer and R.M.~Williams \journal \PL &B273 (91) 438.}

\REF\MIZTAD{S.~Mizoguchi and T.~Tada \journal \PRL &68 (92) 1795.}

\REF\ARC{F.J.~Archer \journal \PL &B295 (92) 199.}

\REF\FELGRA{
G.~Felder and O.~Grandjean, ``On combinatorial three-manifold
invariants'', ETH preprint (1992).}

\REF\MIZONE{S.~Mizoguchi \journal \IJMP &A8 (93) 3909.}

\REF\OOGTHR{H.~Ooguri \journal \PTP &89 (93) 1.}

\REF\TURONE{V.G.~Turaev \journal \CRASP &313 (91) 395;
V.G.~Turaev \journal \JDG &36 (92) 35.}

\REF\OOGONE{H.~Ooguri \journal \NP &B382 (92) 276.}

\REF\CHUFUK{S.~Chung, M.~Fukuma and A.~Shapere, ``Structure of
topological lattice field theories in three dimensions'', preprint
CLNS 93/1200.}

\REF\OOGTWO{H.~Ooguri \journal \MPL &A7 (92) 2799.}

\REF\HOR{A.S.~Schwarz \journal \LMP &2 (78) 247;
G.T.~Horowitz \journal \CMP &125 (89) 417.}

\REF\SASONE{N.~Sasakura \journal \PL &B316 (93) 329. }

\REF\RUS{B.Ye.~Rusakkov \journal \MPL &A5 (90) 693.}

\REF\ALE{J.W.~Alexander \journal \AM &31 (30) 292.}

\REF\GROVAR{M.~Gross and S.~Varsted \journal \NP &B378 (92) 367.}

\REF\VER{E.~Verlinde \journal \NP &B300 (88) 360.}

\REF\MIGTWO{D.V.~Boulatov, V.A.~Kazakov, I.K.~Kostov and A.A.~Migdal
\journal \NP &B275 (86) 641.}

\REF\REVCLU{See, for example, C.~Itzykson and J.M.~Drouffe,
{\it Statistical field theory}, Cambridge (1989), vol.2, chapter 7.}

\REF\NIENAU{B.~Nienhuis and M.~Nauenberg \journal \PRL &35 (75) 477.}

\REF\LGmodel{For review see,
R.~Dijkgraaf, H.~Verlinde and E.~Verlinde,
``Notes on Topological String Theory and 2D Quantum Gravity'',
Trieste Spring School 1990: 91-156; \nextline
R.~Dijkgraaf, ``Intersection Theory, Integrable Hierarchies and
Topological Field Theory'',
NATO ASI: Cargese 1991: 95-158.}

\REF\KAWWAT{N.~Kawamoto and Y.~Watabiki \journal \PR &D45 (92) 605;
N.~Kawamoto and Y.~Watabiki \journal \NP &B396 (93) 326.}

\REF\EFS{See, for example, A.C.D.~van~Enter, R.~Fernandez and
A.D.~Sokal \journal \PRL &66 (91) 3253, and references therein.}


\pubnum{UT-668 \ Revised version}

\titlepage

\title{Discrete Phase Transitions
Associated to Topological Lattice Field Theories in Dimension
$D\geq 2$
}
\author{Naoki Sasakura
\footnote{\star}
{E-mail address: sasakura@tkyux.phys.s.u-tokyo.ac.jp \nextline
Address after April 1, 1994: KEK, Tsukuba, Ibaraki 305, Japan}
\address{Department of Physics, University of Tokyo, Bunkyo-ku,
Tokyo 113, Japan}}

\abstract{
We investigate the neighborhood of Topological Lattice Field
Theories (TLFTs) in the parameter space of general lattice field
theories in dimension $D\geq 2$, and discuss the phase structures
associated to them. We first define a volume-dependent TLFT,
and discuss its decomposition to a direct sum of irreducible
TLFTs, which cannot be decomposed anymore. Using this decomposed
form, we discuss phase structures and renormalization group flows
of volume-dependent TLFTs. We find that TLFTs are on multiple first
order phase transition points as well as on fixed points of the
flow. The phase structures are controlled by the physical states
on $(D-1)$-sphere of TLFTs.
The flow agrees with the Nienhuis-Nauenberg criterion.
We also discuss the neighborhood of a TLFT in general directions
by a perturbative method, so-called cluster expansion.
We investigate especially the $Z_p$ analogue of the Turaev-Viro
model, and find that the TLFT is in general on a higher order
discrete phase transition point. The phase structures depend on
the topology of the base manifold and are controlled by the
physical states on topologically non-trivial surfaces. We also
discuss the correlation lengths of local fluctuations,
and find long-range modes propagating along topological defects.
Thus various discrete phase transitions are associated to TLFTs.
}

\endpage


\chapter{Introduction}

The topological field theory (TFT)\refmark{\WITONE,\ATI}
has been an important tool in
theoretical physics and pure mathematics.
In theoretical physics it has been appearing in the investigations
of the non-perturbative aspects of string theories and quantum
gravities.
The topological phase of two-dimensional quantum gravity was
discussed by Witten\refmark{\WITTWO}.
In three dimensions, he pointed out that Chern-Simons
theories with certain gauge groups are interesting models
for quantum gravity\refmark{\WITTHR}.
The three-dimensional Chern-Simons theories have been also useful
in two-dimensional conformal field theories\refmark{\MOOSEI} and
integrable models\refmark{\WITFOU}.
TFTs are also interesting from mathematical point of view
concerning topological invariants of manifolds\refmark{\WITFIV,
\WITSIX}.

Recently lattice constructions of TFTs have been being discussed.
They are simpler than the usual continuum formulations of TFTs
in the sense that they are free of infinitely many local spurious
degrees of freedom and divergences.
This simpleness is expected to facilitate the systematic
constructions and classifications of TFTs.

In fact, in two dimensions,
topological lattice field theories (TLFTs)
have been constructed and classified
extensively\refmark{\WHEONE-\DURJON}.
It was shown that
the class of TLFTs with degrees of freedom on links
has one-to-one correspondence to semi-simple associative
algebras\refmark{\BACPET-\FUKKAW}.
The relation between this class of TLFTs and known topological
matter field theories in two dimensions\refmark{\MIGONE,\EGUYAN}
was discussed\refmark{\FUKKAW}.
In general, it was shown\refmark{\DURJON}
that, starting with Atiyah's axioms\refmark{\ATI},
any unitary TFT can be decomposed as a direct sum of irreducible
TFTs, which can be realized as TLFTs with degrees of
freedom on vertices.
Here the notion of irreducible TFTs was first introduced, and
they are TFTs which can not be decomposed anymore.
This notion will play an important role in the present paper.

In three dimensions, some systematic constructions of some
series of TLFTs have been done, but any systematic classifications
have not been done yet since they must be tightly  related
to the classification of three-dimensional manifolds.
The first TLFT was constructed by
Ponzano and Regge\refmark{\PONREG},
motivated by the three-dimensional Regge calculus\refmark{\REG}.
The degrees of freedom of their TLFT are the labels of the
irreducible representations of $SU(2)$.
Their TLFT was generalized to a series of
models with the quantum groups $SU_q(2)$ by
Turaev and Viro\refmark{\TURVIR}.
Dijkgraaf and Witten\refmark{\DIJWIT}
constructed models with discrete gauge groups.
A relation between the Turaev-Viro model and
2D conformal field theories was pointed out, and some generalizations
of the model were done\refmark{\DURJAC,\OOGSAS}.
Boulatov\refmark{\BOU}
constructed tensor-model versions of the Turaev-Viro model.
(See also [\ARCWIL]-[\OOGTHR] for other discussions concerning
Turaev-Viro model.)
The relations to continuum formulations have not been
studied enough.
But it was shown by Turaev\refmark{\TURONE} that the Turaev-Viro
model is equivalent to the square of the Chern-Simons theory
with the gauge group $SU(2)$.
Ooguri and the present
author\refmark{\OOGSAS,\OOGONE} discussed the equivalence between
the Ponzano-Regge model and the Chern-Simons theory
with gauge group $ISO(3)$.
A new class of TLFTs was constructed by Chung, Fukuma and
Shapere\refmark{\CHUFUK}.
This class of TLFTs is defined on arbitrary 3D cell complexes,
not only on tetrahedral lattices. These TLFTs were shown to
have one-to-one correspondence with a class of Hopf algebras.

In four dimensions, Ooguri\refmark{\OOGTWO}
generalized the Boulatov's argument
to four dimensions and obtained the 4D analogies
of the Ponzano-Regge-Turaev-Viro model. He argued that
the 4D analogue of the Ponzano-Regge model
is the lattice realization of the $BF$-theory\refmark{\HOR}.

The recent developments on TLFTs have been concentrated only on
constructions and classifications of TLFTs, and seem
to uncover physical content. Since TLFTs are special cases
of general lattice field theories, the investigation of the
positions of TLFTs in the parameter space of general
lattice field theories will clarify the specialties of TFTs
and the background reason of their
appearances in the investigations of quantum gravity and string
theory.
The present author investigated the phase structure around
2D TLFTs by a perturbative scheme and found that they
are in general on multiple first order phase
transition points\refmark{\SASONE}.
A usual continuum field theory describes
the long-range modes near a fixed point of a renormalization
group flow on a continuous
phase transition surface. Thus, complimentary to usual
continuum field theories, TLFTs(TFTs) might be useful
in describing the physics near a fixed point on a
discrete phase transition surface.

This paper is an extension of the present author's previous paper
to dimension $D\geq2$.
The  paper is organized as follows.

In section 2, we will define volume-dependent TLFTs in
any dimension $D(\geq 2)$,
and discuss their decompositions under certain conditions.
We will review the physical structures
of TLFTs\refmark{\TURVIR}.
Then we will decompose a volume-dependent TLFT satisfying
certain conditions into
a direct sum of irreducible TLFTs with volume-dependent
numerical factors.
Using this decomposed form, we will show that
a TLFT is in general
on a multiple first order phase transition point,
which is also a fixed point of the renormalization group flow
in the parameter space of volume-dependent TLFTs.
As an example, the procedure of the decomposition
is shown in the two-dimensional lattice
QCD\refmark{\MIGONE,\RUS,\WHETWO}.

In section 3, we will investigate the neighborhood of a TLFT
by a perturbative method.
We will discuss the method in detail, which is a kind of
cluster expansion.
Using this method, we will investigate the phase structure near
a TLFT in general directions,
and again will find the structure of first order phase transitions
in the first order of the cluster expansion.
The number of the different phases around it is shown to be equal
to the number of the irreducible TLFTs of which the TLFT is a
direct sum.
The higher order contributions do not change this structure
qualitatively, provided the perturbative treatment is valid
around a TLFT and that the base manifold is topologically trivial.
As a simple example, we will discuss the neighborhoods
of the TLFTs in the parameter space of the 2D $Q$-state Potts
model\refmark{\BACPET}.

In section 4, we will investigate the phase structure
near the discrete group $Z_p$ analogue\refmark{\DIJWIT,\BOU}
of the Turaev-Viro model on a topologically non-trivial manifold
in three dimensions.
This model is an irreducible TLFT, hence
is not on a first order phase transition point.
But we will find that the model is in general on a
higher order discrete phase transition point,
which is controlled by the physical states on topologically
non-trivial surface.
To show it, the higher orders of the cluster
expansion are essential.

In section 5, we will give summary, comments and discussions.


\chapter{Volume-dependent topological lattice field theories}

For later discussions, it is more convenient to enlarge the
parameter space of TLFTs. We will
begin with defining volume-dependent TLFTs,
which do not have any local degrees of freedom like TLFTs
but are dependent on the volume of the manifold.

The rough description of the usual constructions of TLFTs is
as follows.
Topological field theories are field theories
which are invariant under
local deformations of background metric.
Physical quantities of such theories depend only on the
topology of the underlying manifold.
In TLFTs,
the independence of the background metric of TFTs corresponds
to the independence of the lattice structure.
Consider a simplicial decomposition of a given manifold.
Usually lattice theories are defined by associating degrees of
freedom to the simplices and assigning local weights.
The partition function is defined as the summation
of the products of the local weights
over the degrees of freedom.
The partition function of a TLFT
must be invariant under the change of
the simplicial decomposition of the manifold.
The simplicial decompositions of a manifold can be generated by
a certain set of local moves in any dimension\refmark{\ALE,
\GROVAR}.
These local moves constrain the local weights. Solving these
constraints, one obtains TLFTs.

In the present paper, we are not interested in concrete models
of TLFTs, but in
the general structures of TLFTs in an arbitrary dimension.
Thus, for convenience, we use a more formal definition
of TLFTs.
Following Atiyah\refmark{\ATI}, we define a TLFT as a set of
partition functions associated to
simplicially decomposed manifolds possibly with boundaries.
The partition functions should be independent of
the simplicial decompositions of the insides of the manifolds.
Hence these partition functions must satisfy certain consistency
conditions under gluing operations of the manifolds.

\noindent
{\it 2.1 Definition of volume-dependent TLFTs}

The lattices we consider in this paper are those
obtained by
simplicial decompositions of $D$-dimensional oriented connected
manifolds possibly with boundaries.
The lattices of boundaries can be colored by associating to each
$l$-dimensional simplex ($l=0,1,\cdots,D-1$) an element $x_l$
of a finite index set $X_l$.

The ingredients of a volume-dependent TLFT are complex-valued
partition functions.
A partition function is associated to a manifold $M$
which is decomposed with $A$
$D$-dimensional simplices and possibly has
colored boundaries $\Sigma(c)$;
$$
Z(M;A;\Sigma(c)).
\eqn\parvtlft
$$
Here $c$ denotes symbolically the coloring.
The partition function is independent of the way how the inside
of the manifold is decomposed with $A$ $D$-dimensional simplices but
depends on the number $A$ and the topology of $M$ as well as
the simplicial decomposition and the coloring of the boundaries.

The partition functions \parvtlft\ must satisfy constraints under
gluing operations of lattices.
Let $(M;A;\Sigma)$ denote a lattice with uncolored boundaries.
Consider a lattice $(M_3;A_3;\Sigma_3)$ obtained by
identifying pairwise some of the $(D-1)$-simplices of the boundaries
of $(M_1;A_1;\Sigma_1)$ and $(M_2;A_2;\Sigma_2)$
(Fig.\figglu ).
Let $b$ denote the sub-boundaries identified pairwise.
Then $A_3=A_1+A_2$, $M_3=M_1\cup_b M_2$ and
$\Sigma_3=\Sigma_1 \cup_b \Sigma_2$, symbolically.
We demand that the partition functions for the three lattices
should satisfy the following constraint under the gluing operation:
$$
Z(M_3;A_3;\Sigma_3(c_3))=\sum_{c_{inn}}
Z(M_1;A_1;\Sigma_1(c_1))g_b^{c_{1b}c_{2b}}Z(M_2;A_2;\Sigma_2(c_2)).
\eqn\glu
$$
Here $g_b^{c_{1b}c_{2b}}$ is the gluing tensor, and
$c_{ib}(i=1,2)$ are the colorings associated to the identified
simplices of the sub-boundaries\foot{
In most of concrete models of TLFTs,
the gluing tensor is the product of the Kronecker's deltas of the
colors of the identified simplices with some numerical factors.}.
The summation is over the colorings associated to
the simplices which become inner simplices after the gluing
operation.

We demand also that the partition functions must satisfy the
following constraint under a self-gluing operation:
$$
Z(M_2;A;\Sigma_2(c_2))=\sum_{c_{inn}}g_b^{c_{1b}c_{1b}'}
Z(M_1;A;\Sigma_1(c_1)).
\eqn\slfglu
$$
Here $(M_2;A;\Sigma_2)$ is obtained by identifying some of the
$(D-1)$-simplices of the boundaries $\Sigma_1$ (Fig.\figslfglu ).

\noindent
{\it 2.2 Laplace decomposition}

In this subsection, we will decompose a volume-dependent TLFT
satisfying a certain condition into
a direct sum of TLFTs by applying an analogue of
the Laplace transformation to
the volume-dependence of the partition functions.
Then, in the following subsections, we will decompose each TLFT
into a direct sum of irreducible TLFTs.

Let ${\cal T}^D$ denote a $D$-dimensional simplex, and
$\pt{\cal T}^D$ its boundary lattice.
We consider only a TLFT satisfying the following condition:
{\it A discrete analogue of the Laplace transformation is
applicable to the volume
dependence of the partition function for a $D$-dimensional
ball $B^D$
with $\pt {\cal T}^D$ as its boundary.}
$$
Z(B^D;A;\pt {\cal T}^D(c))=\sum_i Z_i(B^D;\pt {\cal T}^D(c))
{\omega_i}^A,
\eqn\lts
$$
{\it where $\omega_i$'s are complex numbers and $i$ is the label
to distinguish different $\omega_i$'s.}

First we will prove that the $Z_i(B^D;\pt {\cal T}^D(c))$'s
are orthogonal.
Consider two lattices $(B^D;A_i;\pt {\cal T}^D)\ (i=1,2)$ and
a lattice $(B^D;A_1+A_2;\Sigma)$ obtained by
gluing the two lattices identifying one of the $(D-1)$-simplices
on $\pt{\cal T}^D$ of each lattice (Fig.\figtwosim).
Since the partition functions must
satisfy the constraint \glu\ for the gluing operation, one obtains
$$
Z(B^D;A_1+A_2;\Sigma(c_3))=
\sum_{c_{inn}}
Z(B^D;A_1;\pt {\cal T}^D(c_1))g_{{\cal T}^{D-1}}^{c_{1b}c_{2b}}
Z(B^D;A_2;\pt {\cal T}^D(c_2)).
\eqn\twopar
$$
Using the assumption \lts, the right-hand side is rewritten as
$$
\sum_{i,j}\sum_{c_{inn}}
Z_i(B^D;\pt {\cal T}^D(c_1))g_{{\cal T}^{D-1}}^{c_{1b}c_{2b}}
Z_j(B^D;\pt {\cal T}^D(c_2))
{\omega_i}^{A_1} {\omega_j}^{A_2}.
\eqn\twolap
$$
Since the left-hand side of \twopar\ is a function of $A_1+A_2$,
so must be \twolap. Thus the $Z_i$'s must be orthogonal, and
we obtain the Laplace transformation
for the left-hand side of \twopar\ with
$$
\delta_{ij}Z_i(B^D;\Sigma(c_3))=\sum_{c_{inn}}
Z_i(B^D;\pt {\cal T}^D(c_1))g_{{\cal T}^{D-1}}^{c_{1b}c_{2b}}
Z_j(B^D;\pt {\cal T}^D(c_2)).
\eqn\ort
$$

In the self-gluing operation \slfglu, if the Laplace transformation
is applicable to the right-hand side, then so is the left-hand
side. The $Z_i$'s in the both sides are related by
$$
Z_i(M_2;\Sigma_2(c_2))=\sum_{c_{inn}}g_b^{c_{1b}c_{1b}'}
Z_i(M_1;\Sigma_1(c_1)).
\eqn\slfglud
$$

Any partition function can be obtained by gluing operations
of $D$-simplices and self-gluing operations. Thus
repeating the same discussions as above, we obtain
that the Laplace transformation using the same $\omega_i$'s
is applicable to the volume-dependence of any partition function.
Then the constraint \glu\ is rewritten using $Z_i$'s as
$$
\delta_{ij}Z_i(M_3;\Sigma_3(c_3))=\sum_{c_{inn}}
Z_i(M_1;\Sigma_1(c_1))g_b^{c_{1b}c_{2b}}Z_j(M_2;\Sigma_2(c_2)).
\eqn\glud
$$

The volume-dependence has disappeared in \slfglud\ and \glud.
Since $Z_i$'s are orthogonal for different $i$'s,
the constraints \slfglud\ and \glud\ are
the defining constraints for the partition functions of a TLFT
for each $i$, as will be explained in the following subsection.
Thus we have shown the following.
{\it A volume-dependent TLFT with \lts\ is
a direct sum of TLFTs with some volume-dependent numerical
factors:}
$$
Z(M;A;\Sigma(c))=\sum_{i=1}^N Z_i(M;\Sigma(c)) {\omega_i}^A,
\eqn\abca
$$
{\it where $N$ is a finite number\foot{ \rm
The finiteness can be shown as follows.
In the following subsection, we will discuss the physical Hilbert
spaces of TLFTs and volume-dependent TLFTs. It will be shown that
the dimension of the physical Hilbert space on the $(D-1)$-sphere
is non-zero and finite. Thus $N$ must be finite.},
and $Z_i$'s are the partition functions of TLFTs.}

\noindent
{\it 2.3 Definitions concerning TLFTs}\refmark{\TURVIR}

To prepare to decompose \abca\ further,
we will discuss some elementary definitions concerning TLFTs
in this subsection.

The ingredients of a TLFT are complex-valued partition functions.
A partition function of a TLFT is
associated to a manifold $M$ which is simplicially decomposed
with some $D$-simplices and possibly has colored boundaries:
$$
Z(M;\Sigma(c)).
\eqn\partl
$$
This partition function is independent of the way how the inside
of $M$ is decomposed and must satisfy the constraints \slfglud\
and \glud.

The Hilbert space ${\cal H}_\Sigma$ associated to a
simplicially decomposed
boundary $\Sigma$ is the module freely generated
by all possible colorings of $\Sigma$ over complex numbers.
Thus a wave function in
${\cal H}_\Sigma$ is a complex-valued function of colorings on
$\Sigma$;${\cal H}_\Sigma=\{ \Phi_\Sigma (c_\Sigma)\}$.
The Hilbert space is finite dimensional since the index set $X_l$
is finite.

Consider a $(D-1)$-dimensional closed manifold $\Sigma$,
and $M=\Sigma\times [0,1]$.
Take simplicial decompositions
$\Sigma_1$ and $\Sigma^*_2$
for the two boundaries of $M$
so that one can take a simplicial decomposition of $M$
consistent with $\Sigma_1$ and $\Sigma^*_2$.
Here the symbol $*$ is for the
reverse of orientation. We define a linear map
which maps a state in ${\cal H}_{\Sigma_1}$ to a state in
${\cal H}_{\Sigma_2}$:
$$
\eqalign{
\Phi_{\Sigma_2}(c_2)&=P_{\Sigma_2 {\Sigma}_1}(c_2,c_1)
\Phi_{\Sigma_1}(c_1), \cr
P_{\Sigma_2 {\Sigma}_1}(c_2,c_1)
&=g_{\Sigma_2} ^{c_2 c_2'}Z(\Sigma\times [0,1];
\Sigma^*_2(c_2'),{\Sigma}_1(c_1)),
}
\eqn\pro
$$
where the repeated indices for the colorings are summed over.
{}From now on, it is supposed that the repeated indices for
colorings are summed over unless otherwise stated.

Since the manifold $M=\Sigma\times [0,1]$ can be obtained by
gluing two $M$'s
identifying the boundary $\Sigma$ of one $M$ and $\Sigma^*$ of
another $M$, the linear maps satisfy the following equation:
$$
P_{\Sigma_3 \Sigma_1}(c_3,c_1)=
P_{\Sigma_3 \Sigma_2}(c_3,c_2)
P_{\Sigma_2 \Sigma_1}(c_2,c_1).
\eqn\ppp
$$
This equation \ppp\ is derived from the definition \pro\
and the constraint \glud\ under the gluing operation above.
Taking $\Sigma_3=\Sigma_2=\Sigma_1$, we see that
$P_{\Sigma_1\Sigma_1}=P_{\Sigma_1\Sigma_1}P_{\Sigma_1\Sigma_1}$,
{\it i.e.}, the map $P_{\Sigma_1\Sigma_1}$ is a projection map.

The physical Hilbert space ${\cal H}_{\Sigma_1}^{phys}$ is defined
as the invariant subspace of ${\cal H}_{\Sigma_1}$
by the projection operator $P_{\Sigma_1\Sigma_1}$.
The physical Hilbert space is finite dimensional since the
Hilbert space is finite dimensional.
Take a basis of the physical Hilbert space and its dual basis:
$$
\eqalign{
\phi_{\Sigma_1}^i(c)&\in {\cal H}_{\Sigma_1}^{phys}, \cr
\phi^*_{\Sigma_1\ i}(c)&\in {{\cal H}_{\Sigma_1}^{phys}}^* ,\cr
\phi^*_{\Sigma_1\ i}(c)\phi_{\Sigma_1}^j(c)&=\delta_i^j ,\cr
i&=1,\cdots, {\rm dim}({\cal H}_{\Sigma_1}^{phys}).
}
\eqn\bas
$$
Taking $\Sigma_3=\Sigma_1$ in \ppp,
one finds $P_{\Sigma_1 \Sigma_1}=P_{\Sigma_1 \Sigma_2}
P_{\Sigma_2 \Sigma_1}$. Thus $P_{\Sigma_2 \Sigma_1}$ is
a one-to-one map from ${\cal H}_{\Sigma_1}^{phys}$ to
${\cal H}_{\Sigma_2}^{phys}$.
Using the one-to-one maps $P_{\Sigma_1 \Sigma_2}$ and
$P_{\Sigma_2 \Sigma_1}$, one can
obtain a natural basis and a natural dual basis
for ${\cal H}_{\Sigma_2}^{phys}$:
$$
\eqalign{
\phi_{\Sigma_2}^i(c_2)
&=P_{\Sigma_2 \Sigma_1}(c_2,c_1)\phi_{\Sigma_1}^i(c_1), \cr
\phi^*_{\Sigma_2\ i}(c_2)
&=\phi^*_{\Sigma_1\ i}(c_1)P_{\Sigma_1 \Sigma_2}(c_1,c_2).
}
\eqn\basd
$$
{}From now on, the physical states labeled with the same index are
supposed to be related in this way.

Consider a manifold $M$ with boundaries
$\S^*_1,\cdots,\S^*_n,\S_{n+1},\cdots,\S_m$ with certain simplicial
decompositions.
Following \pro, define
$$
\eqalign{
P&_{M;\S^*_1,\cdots,\S^*_n;\S_{n+1},\cdots,\S_m}
(c_1,\cdots,c_n;c_{n+1},\cdots,c_m)\cr
&=g_{\S_1}^{c_1c_1'}\cdots g_{\S_n}^{c_nc_n'}
Z(M;\S^*_1(c_1'),\cdots,\S^*_n(c_n'),\S_{n+1}(c_{n+1}),\cdots,
\S_m(c_m)),
}
\eqn\prom
$$
which can be regarded as a linear map from
${\cal H}_{\S_{n+1}}\otimes\cdots\otimes{\cal H}_{\S_m}$ to
${\cal H}_{\S_{1}}\otimes\cdots\otimes{\cal H}_{\S_n}$.
Since gluing manifolds $\S_i\times
[0,1]\ (i=1,\cdots, m)$
to $M$ by identifying $\S_i$'s and $\S_i^*$'s does not change the
manifold $M$,
one can show, using the definitions
\pro\ and \prom\ and the constraint \glud\ under the gluing
operation, that
$$
\eqalign{
&P_{M;\S^*_1,\cdots,\S^*_n;\S_{n+1},\cdots,\S_m}
(c_1,\cdots,c_n;c_{n+1},\cdots,c_m)\cr
&=P_{\S_1 \S_1'}(c_1,c_1')\cdots P_{\S_n \S_n'}(c_n,c_n')
P_{M;{\S_1'}^*,\cdots,{\S_n'}^*;\S_{n+1}',\cdots,\S_m'}
(c_1',\cdots,c_n';c_{n+1}',\cdots,c_m')\cr
&~~~~~~~~\times P_{\S_{n+1}' \S_{n+1}}(c_{n+1}',c_{n+1})
\cdots P_{\S_m' \S_m}(c_m',c_m),}
\eqn\pppp
$$
where $\S_i$ and $\S_i'$ denote the same boundary possibly with
different simplicial decompositions.
The equation \pppp\ is a generalization of \ppp.
Thus, taking $\S_i=\S_i'$ in
\pppp, the linear map \prom\
is in fact a map among the physical Hilbert spaces;
${\cal H}_{\S_{n+1}}^{phys}\otimes\cdots\otimes
{\cal H}_{\S_m}^{phys}$ to
${\cal H}_{\S_{1}}^{phys}\otimes\cdots\otimes{\cal H}_{\S_n}^{phys}$.
This fact leads us naturally
to the following definition of a physical correlation function:
$$
\eqalign{
Z&(M;(\S_1)_{i_1},\cdots,(\S_n)_{i_n},(\S_{n+1})^{i_{n+1}},\cdots,
(\S_m)^{i_m})\cr
&\equiv
\phi_{\S_1\ i_1}^*(c_1)\cdots \phi_{\S_n\ i_n}^*(c_n)
P_{M;\S^*_1,\cdots,\S^*_n;\S_{n+1},\cdots,\S_m}
(c_1,\cdots,c_n;c_{n+1},\cdots,c_m)\cr
&~~~~~~~~~\times
\phi_{\S_{n+1}}^{i_{n+1}}(c_{n+1})\cdots \phi_{\S_m}^{i_m}(c_m).
}
\eqn\phycor
$$
The remarkable property of a physical correlation function is that
it is
independent of the simplicial decompositions of its boundaries.
This fact can be proven
by using the definition of the basis of the physical
Hilbert space
\basd\ and \pppp.
In particular, for any $\S$,
$$
Z(\S\times [0,1];(\S)_i,(\S)^j)=\delta_i^j,
\eqn\triort
$$
because the projection map \pro\ is the identity on
${\cal H}_\S^{phys}$.

\noindent
{\it 2.4 Decomposition of a TLFT}

Here we will discuss the decomposition of each TLFT in
\abca\ into irreducible TLFTs.
An irreducible TLFT is defined as a TLFT
which cannot be decomposed further as a direct sum of TLFTs.
This is an extension of the work by
Durhuus and J{\`o}nsson\refmark{\DURJON}, where
they discussed the decomposition of a unitary two-dimensional TFT
into a direct sum of irreducible TFTs.

The discussions will be roughly as follows.
First we will discuss the correlation functions of $S^{D-1}$s,
and will find a convenient basis of the physical Hilbert
space on $S^{D-1}$, which we call the canonical basis.
This basis was first introduced by Verlinde\refmark{\VER}
in the discussions of fusion rules of 2D conformal field theories,
and was applied to 2D TLFTs in [\BACPET] - [\FUKKAW] and [\DURJON].
Then we will show that any partition(correlation)
functions of a TLFT can be expressed as a sum of
the partition(correlation) functions labeled by the labels
of the canonical basis and that these labeled partition functions
are orthogonal.

Here we begin with the notations to be used.
$B_0$ denotes $S^{D-1}$
with a certain simplicial decomposition.
Since, in the physical correlation functions,
the simplicial decompositions of boundaries are irrelevant,
we use $\phi^i$ for the physical state $\phi_{B_0}^i$ and
$\phi_i$ for $\phi_{B_0\ i}$.
We also use $(\Sigma)^i(c)$
for $\phi_{\Sigma}^i(c)$ and $(\Sigma)_i(c)$
for $\phi^*_{\Sigma\ i}(c)$, for short.

Define the following symmetric tensors which reverse the orientations
of the physical states on $S^{D-1}$:
$$
\eqalign{
\eta_{ij}&=Z(T_{S^{D-1}};\phi_i,\phi_j)\cr
&=(B_0)_i(c_1)g_{B_0}^{c_1c_2} Z(T_{S^{D-1}};
B_0^*(c_2),B_0^*(c_3))g^{c_3c_4}(B_0)_j(c_4),\cr
\eta^{ij}&=Z(T_{S^{D-1}};\phi^i,\phi^j),
}
\eqn\metten
$$
where $T_{S^{D-1}}=S^{D-1}\times [0,1]$.
Since the projection operator \pro\ is the identity on the physical
Hilbert space, the following equation holds for any $\S$:
$$
\sum_{i} (\S)^i(c_1)(\S)_i(c_2)=P_{\S \S}(c_1,c_2),
\eqn\prophi
$$
{}From now on it is supposed
that contracted indices labeling the physical states
are summed over unless otherwise stated.
Using \prophi, \triort\ and \glud\ for the operation that
two $T_{S^{D-1}}$s are glued to make one
$T_{S^{D-1}}$, we obtain
$$
\eta_{ij}\eta^{jk}=\delta_i^k,
\eqn\abcb
$$
which shows that the tensors \metten\ are not degenerate.
These tensors can be used to change the cases of the indices
of $\phi_i$ and $\phi^i$.
The proof is as follows:
$$
\eqalign{
Z(M;\phi^i,\S(c))&=Z(T_{S^{D-1}};\phi^i,B_0(c_1))g_{B_0}^{c_1c_2}
Z(M;B_0^*(c_2),\S(c))\cr
&=Z(T_{S^{D-1}};\phi^i,B_0(c_1))P_{B_0 B_0}(c_1,c_2)g_{B_0}^{c_2c_3}
Z(M;B_0^*(c_3),\S(c))\cr
&=Z(T_{S^{D-1}};\phi^i,B_0(c_1))\phi^j(c_1)
\phi_j(c_2)g_{B_0}^{c_2c_3}
Z(M;B_0^*(c_3),\S(c))\cr
&=\eta^{ij}Z(M;\phi_j,\S(c)),
}
\eqn\cas
$$
where we have used the invariance of the topology of $M$ under
gluing a $T_{S^{D-1}}$
to $M$ and the equations \glud, \metten\ and
\prophi.

Next we will obtain the fusion rule of the states on $S^{D-1}$.
Define the following rank-three symmetric tensor:
$$
\eqalign{
N^{ijk}&=Z(S^D;\phi^i,\phi^j,\phi^k),\cr
}
\eqn\thr
$$
where the manifold truly treated can be uniquely obtained by
taking away 3 $D$-simplices from $S^D$.
{}From now on, we will use such implications for simplicity.
Using this tensor \thr, one can reduce the number of $S^{D-1}$s
in correlation functions:
$$
\eqalign{
Z(M;\S(c),\phi^i){N_i}^{jk}&=Z(M;\S(c),B_0(c_1))(B_0)^i(c_1)
(B_0)_i(c_2)g_{B_0}^{c_2c_3}Z(S^D;B^*_0(c_3),\phi^j,\phi^k) \cr
&=Z(M;\S(c),B_0(c_1))P_{B_0B_0}(c_1,c_2)
g_{B_0}^{c_2c_3}Z(S^D;B^*_0(c_3),\phi^j,\phi^k) \cr
&=Z(M;\S(c),B_0(c_1))
g_{B_0}^{c_1c_2}Z(S^D;B^*_0(c_2),\phi^j,\phi^k) \cr
&=Z(M;\S(c),\phi^j,\phi^k).
}
\eqn\imp
$$
Here, from the third to the last line,
we glued the manifold obtained by taking away one
$D$-simplex from $S^D$ and the manifold obtained by taking away
one $D$-simplex from $M$. Since the former manifold is
topologically one
$D$-simplex, the resultant manifold is simply $M$.
In this derivation, we have used
the constraint \glud\ for the gluing operation, \pppp, \phycor,
\prophi\ and \thr.
Taking particularly the three-sphere function as the partition
function in the left-hand side of \imp, one obtains
$$
{N_i}^{jk}{N_k}^{lm}={N_i}^{jlm}={N_i}^{lk}{N_k}^{jm},
\eqn\fus
$$
where ${N_i}^{jlm}$ is the four-sphere function defined as same as
\thr.

By the similar discussions as to derive \imp, we obtain a
formula
$$
Z(M;\S(c),\phi^i)Z(S^D;\phi_i)=Z(M;\S(c)).
\eqn\impd
$$
This formula implies also that, if the TLFT is not null, the
dimension of the physical Hilbert space on $S^{D-1}$ must be
non-zero.
Substituting the first partition function of the left-hand side
with a three-sphere function, we obtain
$$
\eqalign{
N^{ijk} Z(S^D;\phi_k)
&=Z(S^D;\phi^i,\phi^j)\cr
&=\eta^{ij},
}
\eqn\onepoi
$$
where we have used the fact that $S^D$ with two holes is
topologically equivalent to $T_{S^{D-1}}$ and \triort.

{}From \imp, \fus\ and \onepoi, one can see that the physical
states $\phi^i$s form a commutative algebra with an identity.
This algebra might be decomposed into mutually annihilating
sub-algebras, and the decomposition will give the decomposition of
the TLFT.
One can discuss the decomposition of the algebra in general,
but in the present paper
we restrict our discussions
only to the case that {\it the algebra has no
radical}\foot{We are not sure whether a TLFT can have
an algebra with a radical.
In two dimensions, no such TLFTs are
known\refmark{\WHEONE-\FUKKAW}.
One can realize such cases only by taking a certain infinite
limit\refmark{\FUKKAW}.}.
This is because, if a radical exists, one will have to
discuss a wider class of TLFTs than those with \lts.
Let $\phi_R$ be a radical element, and consider a volume-dependent
TLFT
$Z_{\delta r}(M;A;\Sigma(c))\equiv
Z(M;\Sigma(c),\exp(A\delta r \phi_R))$.
Since $\phi_R$ is a radical element and a physical state,
the $Z_{\delta r}$ is a polynomial function of $A$.
Thus, to discuss the neighborhood of a TLFT with radical,
it is not sufficient to consider only the volume-dependence of \lts.

Since an algebra without radical can be decomposed into a direct
sum of simple sub-algebras, the commutative algebra can be
simultaneously diagonalized.
We call the following the canonical
basis\refmark{\BACPET-\FUKKAW,\DURJON}:
$$
\eta^{ij}=\delta^{ij},\ \eta_{ij}=\delta_{ij},
\ N^{ijk}=\lambda^i \delta^{ij} \delta^{ik}\ \
(\lambda^i\not= 0),
\eqn\canbas
$$
which is supposed to be used from now on.
{}From \imp\ and \canbas, we obtain the fusion rule:
$$
\phi^i \phi^j \sim \delta^{ij} \lambda^i \phi^i.
\eqn\formone
$$

One can see in \onepoi\ that
the one-sphere function is the inverse of the coefficient
of the three-sphere function:
$$
Z(S^D;\phi^i)={1\over \lambda^i}.
\eqn\onesph
$$
{}From \impd\ and \onesph,
we are lead to the fact that
any correlation(partition) function can be expressed in a sum
$$
\eqalign{
Z(M;\S(c))&=\sum_{i=1}^N{1\over \lambda^i} Z^i(M;\S(c)),\cr
Z^i(M;\S(c))&\equiv Z(M;\S(c),\phi^i),
}
\eqn\decomp
$$
where $N$ is the dimension of the physical Hilbert space on
$S^{D-1}$.
In other words, \decomp\ is
$$
\sum_{i=1}^N {1\over \lambda^i} \phi^i \sim 1.
\eqn\formtwo
$$
{}From now on, to avoid confusions, any repeated indices of
physical states without
the summation symbol are not supposed to be summed over.

To show that \decomp\ is a direct sum, we will
prove the orthogonality of $Z^i$s.
Suppose $M_3$ is obtained
by gluing $M_1$ and $M_2$. Then, by using \glud,
\formone\ and \decomp,
$$
\eqalign{
\sum_{c_{inn}}
{1\over \lambda^i}Z^i(M_1;\S_1(c_1))g_b^{c_{1b}c_{2b}}
{1\over \lambda^j}Z^j(M_2;\S_2(c_2))
&={1\over \lambda^i\lambda^j}Z(M_3;\S_3(c_3),\phi^i,\phi^j) \cr
&=\delta^{ij} {1\over \lambda^i} Z^i(M_3;\S_3(c_3)).
}
\eqn\canort
$$

We will next discuss the decompositions of the physical Hilbert space
and the physical correlation(partition) functions.
The projection operator specifying the physical Hilbert space
can be decomposed into
$$
\eqalign{
P_{\S \S}&=\sum_i P^i_{\S \S}\cr
P^i_{\S \S}P^j_{\S \S}&=\delta^{ij}P^i_{\S \S}\cr
P^i_{\S \S}(c_2,c_1)
&\equiv {1\over \lambda^i} g_{\S} ^{c_2 c_2'} Z^i(\S\times [0,1];
{\Sigma^*}(c_2'),{\Sigma}(c_1)),
}
\eqn\decpro
$$
where we have used \pro, \decomp\ and
the orthogonality \canort.
Define the $i$-th physical Hilbert space ${\cal H}_\S^{phys,i}$
as the invariant subspace of the projection operator $P^i_{\S \S}$.
Then \decpro\ implies that ${\cal H}_\S^{phys}=\sum_i \oplus
{\cal H}_\S^{phys,i}$.
In particular, the dimension of
each physical Hilbert space on $S^{D-1}$ is one:
dim(${\cal H}_{S^{D-1}}^{phys, i}$)=1.

The physical states in the physical Hilbert spaces
labeled by different $i$ and $j$,
${\cal H}_{\S}^{phys, i}$ and ${\cal H}_{\S'}^{phys, j}$ $(i\neq j)$,
can not interact with each other.
This is because the labeled partition(correlation) function
is orthogonal to the  physical Hilbert space with a different label:
$$
\eqalign{
Z^i(M;\S_1(c_1),(\S_2)^j)
&=Z^i(M;\S_1(c_1),\S_2(c_2))P^j_{\S_2 \S_2}(c_2, c_3)
(\S_2)^j(c_3)\cr
&=\delta^{ij} Z^i(M;\S_1(c_1),(\S_2)^j),\cr
}
\eqn\ortparphy
$$
where $(\S_2)^j\in {\cal H}_{\S_2}^{phys, j}$, and,
in this derivation,
we have used \pppp, \decpro\ and the orthogonality \canort.
The discussion is similar for the case of the dual states.
Thus the linear map defined with the partition function $Z$ in \prom\
is a direct sum of the linear maps defined with
${1\over \lambda^i}Z^i$s, which are the maps from
${\cal H}_{\S_{n+1}}^{phys,i}\otimes\cdots\otimes
{\cal H}_{\S_m}^{phys,i}$ to
${\cal H}_{\S_{1}}^{phys,i}\otimes\cdots
\otimes{\cal H}_{\S_n}^{phys,i}$, respectively.
This concludes that the original TLFT is a direct sum of
the theories whose partition functions are given by
${1\over \lambda^i}Z^i$s.

The orthogonality \canort\ implies that
${1 \over \lambda^i}Z^i$s of each $i$
satisfy the defining constraints
of a TLFT \slfglud\ and \glud.
Thus ${1\over \lambda^i} Z^i(M;\S(c))$s of each $i$
define a TLFT. Since the physical Hilbert
space on $S^{D-1}$ of each TLFT is one-dimensional,
each TLFT is an irreducible TLFT, which cannot be decomposed anymore
into a direct sum of TLFTs.

In the present paper we do not discuss the general case of the algebra,
but one will be able to extend the discussions so far to the
general case.
An irreducible TLFT will have the algebra which cannot be decomposed
anymore into a direct sum of mutually annihilating sub-algebras.

Decomposing the TLFTs (without radical)
in \abca\ into a direct sum of irreducible TLFTs,
a partition function of a volume-dependent TLFT is a direct sum
of those of irreducible TLFTs with volume-dependent
numerical factors:
$$
Z(M;A;\S(c))=\sum_{i=1}^N Z_i(M;\S(c)) (\omega_i)^A,
\eqn\volirr
$$
where the $Z_i$ has only one physical state on $S^{D-1}$.

Since a volume-dependent TLFT with \lts\ is a direct sum
of irreducible TLFTs,
the natural definition of a physical Hilbert space
of a volume-dependent TLFT is the
direct sum of the physical Hilbert spaces of the irreducible TLFTs:
${\cal H}_\S^{phys}=\sum_{i=1}^N \oplus {\cal H}_\S^{phys,i}$.
In fact, a partition function of
a volume-dependent TLFT can be regarded as a multi-linear
map among ${\cal H}_\S^{phys}$ as same as the map \prom\ of
a TLFT.
The physical correlation function defined
similarly to that of TLFTs is independent of the simplicial
decomposition of $M$ and its boundary,
but is dependent on the volume of the manifold $M$.

\noindent
{\it 2.5 Renormalization group flow}

Since a physical correlation
function of a volume-dependent TLFT is dependent on the volume,
there is a renormalization group flow.
A renormalization group flow of a lattice theory
is determined by how the change
of the lattice structure is absorbed in the change of the
parameters of the theory  without changing the physical outcomes.
Since a physical correlation function of a volume-dependent
TLFT depends only on the
volume $A$ of the lattice $M$, the invariance of the physical
outcomes imposes only that
$(\omega_i)^A=(\omega_i')^{A'}$
for each $i$ under the change of the volume from $A$ to $A'$.
Taking the extensive variables $p_i=\ln (\omega_i)$
as the parameters of the theory, the flow is $p_iA=p_i'A'$.
This implies the volume-dimension of the parameter
$p_i$ is 1, and this agrees with
the Nienhuis-Nauenberg criterion\refmark{\NIENAU}.

The fixed points of the renormalization group flow are the points
with $\omega_i=0,1$(or $p_i=-\infty,0$) for all $i$.
The point $p_i=-\infty$ is the infrared fixed point,
and the point $p_i=0$ is the ultraviolet fixed point.
Since the volume-dependence vanishes on such a fixed point,
the theory becomes a TLFT.
On the other hand, one can make a volume-dependent TLFT
from a TLFT by \volirr\ with certain $\omega_i$s.
The original TLFT corresponds to the point $\omega_i=1$ for all $i$.
Hence a TLFT can be regarded as being on a
fixed point in the parameter space of a
volume-dependent TLFT, and this fixed point has
relevant operators with volume-dimension one.
The number of such operators
is equal to the dimension of the physical Hilbert space on $S^{D-1}$
of the TLFT.

\noindent
{\it 2.6 First order phase transitions and order parameters}

We will consider the thermo-dynamical limit $A\rightarrow\infty$.
{}From \volirr, a partition function of a volume-dependent
TLFT is
$$
Z(M;A)=\sum_{i=1}^N Z_i(M) \exp (p_iA),
\eqn\vdtlftpar
$$
where $M$ is without any boundaries and $N$ is the number of
the irreducible TLFTs.
The free energy per $D$-simplex in the thermo-dynamical limit is
$$
f=-\lim_{A\rightarrow\infty} {1\over A} \ln Z=-{\rm max}_{i}(p_i),
\eqn\freenergy
$$
where ${\rm max}_{i}(p_i)$ is for taking the $p_i$ which has the
largest real part among all the $p_i$s.
Thus the point $p_i=0$ for all $i$ is a multiple first order phase
transition point with $N$ different phases around it
\foot{ In general, $Z_i(M)$
might be zero, and then the discussions must be
changed. Here we set aside such special cases.}.
In general, a TLFT is on a multiple first order phase
transition point, and the number of the phases around it
is equal to the number of the irreducible TLFTs of which
the TLFT is the direct sum.

The order parameters
are appropriately given by using the one-sphere functions.
Define the one-sphere expectation values as follows:
$$
\langle \phi^i \rangle^M =\lim_{A\rightarrow \infty}
{Z(M;A;\phi^i)\over Z(M;A)},
\eqn\expdef
$$
where $\phi^i$ is a physical state on $S^{D-1}$, and we take
the canonical basis to label them.
Using the orthogonality \ortparphy, we obtain
$$
{1\over \lambda^i} \langle \phi^i \rangle^M=
\lim_{A\rightarrow \infty} {1\over \lambda^i}
{Z_i(M;\phi^i)\exp(p_iA) \over
\sum_{j=1}^N Z_j(M) \exp (p_jA)}=\delta^{im},
\eqn\expnum
$$
where we assumed the system is in the $m$-th phase, that is,
$p_m$ has the greatest real part among $p_i(i=1,\cdots,N)$.
Thus the one-sphere expectation values give good order parameters.

Comments are in order. So far
we have assumed implicitly that there are no constraints on the
parameters of the theories and that we can take any values
for the parameters $\omega_i$s or $p_i$s.
But in some physical cases we cannot.
An example is the 2-dimensional lattice QCD in the following
subsection.
This theory is a volume(area)-dependent TLFT,
but cannot have
any first order phase transition points if the positivity
of the local weight is assumed.

\noindent
{\it 2.7 An example --- 2D lattice QCD}

As a simple but physically interesting example of a
volume(area)-dependent TLFT, we consider the 2-dimensional
lattice QCD\refmark{\MIGONE,\RUS,\WHETWO,\FUKKAW}.
As the index set $X_1$ for 1-simplex(edge) we take the Lie group
$G=SU(n),U(n)$.
The index set is infinite for this model, but nonetheless
we can treat it straightforwardly
in the same way as in the preceding subsections.
The index set for 0-simplex(vertex)
is not considered in this model.
The partition function for a single 2-simplex(triangle)
and the gluing tensor are defined as
$$
\eqalign{
Z({\cal T}^2;1;\pt {\cal T}^2 (g_1,g_2,g_3))
&=C_{g_1,g_2,g_3}=z(g_1g_2g_3),\cr
z(ugu^{-1})&=z(g), \cr
g^{g_1g_2}&=\delta(g_1,g_2^{-1}), \cr
}
\eqn\locwei
$$
where $u,g,g_i \in G$, and ${\cal T}^2$ is a 2-simplex(triangle).
We take the local weight
$z(g)$ as a positive real function of the group
element of $G$, and it satisfies the invariance
in the second line.
The $g^{g_1g_2}$ is the gluing tensor for a pair of 1-simplices,
and is expressed by the delta function associated to
the Haar measure of $G$, which is the measure used in summing over
the internal colorings.

To show that the elementary data \locwei\ define the partition
functions of a volume-dependent TLFT, the following
invariance of the product of two local weights
under the flip move (Fig.\figflip) is essential:
$$
\eqalign{
\int dg z(g_1g_2g)z(g^{-1}g_3g_4)
&=\int d(g_2^{-1}gg_4)z(g_1g_2(g_2^{-1}gg_4))z((g_2^{-1}gg_4)^{-1}
g_3g_4) \cr
&=\int dg z(g_1gg_4)z(g^{-1}g_2g_3). \cr}
\eqn\flipinv
$$
It is known that all triangulations of a
two-dimensional surface with the same number of triangles
can be generated by the flip move\refmark{\MIGTWO}.
A partition function $Z(M;A;\S(c))$ can be obtained from the
elementary data \locwei\ by using the rules \glu\ and \slfglu.
The invariance \flipinv\
guarantees that the obtained partition function
is in fact independent of the triangulation of $M$
and that the partition functions satisfy \glu\ and \slfglu.

To perform the Laplace decomposition of the area-dependent TLFT,
we will calculate $Z(B^2;A;\pt {\cal T}^2 (g_1,g_2,g_3))$,
where $B^2$ is a two-dimensional ball.
The invariance in the second line of \locwei\ enables one
to expand the local weight $z$ in the characters of $G$:
$$
z(g)=\sum _R \Lambda_R d_R \chi_R(g),
\eqn\chrdec
$$
where $d_R$ is the dimension of the irreducible representation $R$
of $G$, and $\Lambda_R$ is the coefficient of the expansion.
The partition function $Z(B^2;A;\pt {\cal T}^2 (g_1,g_2,g_3))$
can be calculated easily using this expanded form \chrdec\ and the
orthogonality of the characters:
$$
Z(B^2;A;\pt {\cal T}^2 (g_1,g_2,g_3))=\sum_R (\Lambda_R)^A d_R
\chi_R(g_1g_2g_3).
\eqn\parbal
$$
This result \parbal\ and the orthogonality of the characters
imply that each $d_R \chi_R$ gives the local weight of a TLFT.
To see that each weight defines an irreducible TLFT, we will check
dim$({\cal H}_{S^1}^{phys\ R})=1$.
We take a three-gon as the simplicially decomposed $S^1$.
The projection operator from a three-gon to another
is calculated to be
$$
\eqalign{
{{P_R}^{g_1,g_2,g_3}}_{g_1',g_2',g_3'}
&=\int dg_4dg_5
{{C_R}^{(g_1g_2g_3)}}_{g_4,g_5}{{C_R}^{g_4,g_5}}_{(g_1'g_2'g_3')}\cr
&=\chi_R((g_1g_2g_3)^{-1}) \chi_R(g_1'g_2'g_3'),\cr
{C_R}_{g_1,g_2,g_3}&\equiv d_R \chi_R(g_1g_2g_3).
}
\eqn\proqcd
$$
This projection operator projects any wave-function
$\phi(g_1,g_2,g_3)$ to $\chi_R((g_1g_2g_3)^{-1})$.
Thus ${\cal H}_{S^1}^{phys\ R}
=\{\chi_R((g_1g_2g_3)^{-1})\}$ and its dimension is one.

Since the 2D lattice QCD
is an area-dependent TLFT with
the decomposition as a direct sum of an infinite number of
TLFTs, one might expect that it has first order phase
transition points. But, in fact, it is not valid because
the coefficient $\Lambda_0$ for the trivial representation
in \chrdec\
has always the greatest absolute value. This inequality can be shown
as the following. Since the local weight $z$ is assumed to be
positive real,
$$
|\Lambda_R|=|{1\over d_R}\int dg \chi_R(g) z(g)|
=|{1\over d_R}\int dg \sum_{i=1}^{d_R} \rho_i(g) z(g)|
< \int dg z(g)=\Lambda_0,
\eqn\lazgre
$$
where $\rho_i$s are the eigenvalues in the representation $R$.


\chapter{Neighborhood of topological lattice field theories}

In the previous section we investigated volume-dependent
TLFTs and showed that TLFTs are on fixed points of
the renormalization group flow as well as on
multiple first order phase transition points in the parameter
space of volume-dependent TLFTs.
In this section, we will consider general perturbative deformations
of TLFTs.
To treat such deformed systems,
we will introduce a perturbative method,
which is a kind of cluster expansion\refmark{\REVCLU}.
We assume that
such a perturbative method is available
in the neighborhood of a TLFT.

Consider a manifold $M$
simplicially decomposed with a certain number of $D$-simplices.
The partition function of $M$ of a TLFT is expressed as a
product of local partition functions of the TLFT:
$$
\eqalign{
Z(M)&= \sum_{c} \prod_q Z({\cal T}^D(q);c)
\prod_b g_b^{cc'},\cr
Z({\cal T}^D(q);c)&\equiv Z({\cal T}^D(q);\pt {\cal T}^D(c)),
}
\eqn\parlocrep
$$
where the summation is over the inner colorings, and the products
are of the local partition functions over all the $D$-simplices
and the gluing tensors.

We consider general changes of the local partition functions;
$Z({\cal T}^D(q);c)\rightarrow
Z({\cal T}^D(q);c)+\delta Z({\cal T}^D(q);c).$
We assume that $\delta Z$s are small enough to justify the
perturbative treatment in the following subsection.
We will not consider
the changes in the gluing tensors for simplicity,
because they will not change the discussions essentially.
Since we consider general changes, the deformed system is not a TLFT
anymore. Thus the partition function depends on the
simplicial decomposition of the manifold. Thus, from now on, we
are supposed to consider certain simplicial decompositions and
will take the thermo-dynamical limit that the total number of the
$D$-simplices in $M$ is taken to infinity.

\noindent
{\it 3.1 Cluster expansion}

Here we will introduce a systematic perturbative
method, which
approximates the partition function and the correlation functions
of the deformed system.
This method is a kind of cluster expansion\refmark{\REVCLU},
but is unusual in the sense that it is  a cluster expansion
of an operator.

Here we begin with the partition function.
The partition function of the deformed system is defined by
$$
Z(M;\delta Z)= \sum_{c} \prod_q
(Z({\cal T}^D(q);c)+\delta Z({\cal T}^D(q);c))
\prod_b g_b^{cc'},
\eqn\parper
$$
where $q$ is the label for the $D$-simplices in $M$.
Expanding this partition function \parper\ in $\delta Z$,
we obtain
$$
\eqalign{
Z(M;\delta Z)&=\sum_{n=0}^A {1\over n!}
\sum_{q_1,\cdots, q_n} u_n(q_1,\cdots,q_n),
\cr
u_0&=Z(M),
}
\eqn\exppar
$$
where $A$ is the total number of the $D$-simplices in $M$,
and $Z(M)$ is the partition function of the original TLFT.
The
$u_n(q_1,\cdots,q_n)$ describes an $O((\delta Z)^n)$ contribution
to the partition function, and the $q_i$s specify
the $n$ $\delta Z$s coming from  the $D$-simplices $q_1,\cdots,q_n$.
The $u_n(q_1,\cdots,q_n)$ is supposed to be non-zero only when
$q_1,\cdots,q_n$ are different from each other.

If we assume that the simplicially decomposed manifold $M$
has an enough `extension',
the lower order contributions can be
calculated using the correlation functions of the original
TLFT on $M$:
$$
u_n(q_1,\cdots,q_n)=Z(M;U_n(q_1,\cdots,q_n))
+\theta (n-n(M))Z(M_1;U'_n(q_1,\cdots,q_n))+\cdots,
\eqn\parope
$$
where $n(M)$ is the number of $D$-simplices below which
one does not have to
consider correlation functions on manifolds different from $M$,
and $\theta(n)$ is the discrete analogue of
$\theta$-function: $1$ for $n\geq 0$ and $0$ for $n<0$.
An example of $M$ and $q_1,\cdots,q_n$
that one has to consider correlation
functions on manifolds different from $M$ are given in
Fig.\figdifman.
$U_0$ is the identity and
$U_n$ takes non-zero values only when the $q_i$s are
different from each other.
Since $n(M)=1$ in $D=1$, we will consider only the $D>1$ cases.
The $D=1$ case can be treated very easily in another way.
We assume that the thermo-dynamical limit is taken by
refining the lattice structure of $M$ almost uniformly.
Then $n(M)\rightarrow\infty$ in the thermo-dynamical limit,
so that the right-hand side of \parope\ will be well
approximated by the first term.
Thus we will consider only the first term from now on.

The definition of the first term of \parope\ is as follows.
Consider a lattice $M'$ obtained by taking away $n$
$D$-simplices $q_1,\cdots,q_n$ from $M$.
Then $M'$ is a lattice with some boundaries $\S$, and
a partition function $Z(M';\S(c))$ of the original TLFT
can be associated to the lattice $M'$. Contracting the
colorings of $Z(M';\S(c))$ with those of
$\delta Z({\cal T}^D(q_i);c)$s,
one obtains $Z(M;U_n(q_1,\cdots,q_n))$.
Thus $U_n(q_1,\cdots,q_n)$ represents the operation to take away
$n$ $D$-simplices $q_1,\cdots,q_n$ from $M$ and contract the
colorings of the boundaries with those of
$\delta Z({\cal T}^D(q_i);c)$s.

If the $n$ $D$-simplices $q_1,\cdots,q_n$ are separated,
$Z(M;U_n(q_1,\cdots,q_n))$ is given by contracting the
colorings of the $n$-sphere function of the original TLFT
with those of $\delta Z({\cal T}^D(q_i);c)$s.
Since the $n$-sphere function is independent of the positions $q_i$,
the $q_i$ dependences of $Z(M;$ $U_n(q_1,\cdots,q_n))$
come only from those of $\delta Z({\cal T}^D(q_i);c)$s.
This property simplifies the systematic
calculation of \exppar\ drastically,
because the leading contributions come from the cases
that the $q_i$s are separated.

To consider sub-leading contributions,
we introduce new operations $W_n(q_1,\cdots,q_n)$
satisfying
$$
U_n(q_1,\cdots,q_n)=\sum_{k_1,\cdots,k_n=0
\atop \sum_{i=1}^n ik_i=n}
\left( \prod_{i=1}^n {1\over k_i!(i!)^{k_i}} \right)
\sum_{\sigma_n}
\overbrace{W_1(q_{\sigma_n(1)})\cdots
W_1(q_{\sigma_n(k_1)})}^{k_1\ times}\cdots
\overbrace{W_n(\cdots,q_{\sigma_n(n)})}^{k_n\ times},
\eqn\wdef
$$
where $\sigma_n$ denotes
the permutation maps $\{1,\cdots,n\}\rightarrow
\{1,\cdots,n\}$. The relation \wdef\ defines $W_n$ iteratively
with $U_n$ and $W_1 \sim W_{n-1}$. The relation between $W_n$s
and $Z(M;\delta Z)$ is given by
$$
\eqalign{
Z(M;\delta Z)&=Z(M;\exp(W)),\cr
W&=\sum_{n=1} \sum_{q_1,\cdots,q_n} W_n(q_1,\cdots,q_n),
}
\eqn\expwpar
$$
which can be proven formally using the definitions
\exppar, \parope\ and \wdef.

The relation \wdef\ is of course meaningless before the products
of $W_n$s are defined.
{}From \wdef\ one obtains $W_1(q)=U_1(q)$ without any such products.
But to obtain $W_2$ we have to define the product
$W_1(q_1)W_1(q_2)$.
In fact there are various ways to define the product.
If $q_1$ and $q_2$ are separated, a natural definition of the
product would be to identify $W_1(q_1)W_1(q_2)$ with $U_2(q_1,q_2)$,
because the former can be regarded as
the operation of taking away the $D$-simplices
$q_1$ and $q_2$.
But how about the cases that $q_1$ and $q_2$ coincide or are
next to each other?
We use `operator splitting regularization' for the definitions
of the products in general.
The meaning is as follows.
For example, $Z(M;W_1(q_1)W_1(q_2))$ is
defined by contracting the
colorings of $\delta Z({\cal T}^D(q_i);c_i)\ (i=1,2)$ with those of
the two-sphere function $Z(M;\pt{\cal T}^D(c_1),\pt{\cal T}^D(c_2))$
of the original TLFT.
This definition
coincides with $U_2(q_1,q_2)$ if $q_1$ and $q_2$ are separated.
{}From now on, we assume that
such `operator splitting regularization' is used
as the definitions of the products of $W_n$s.

Since $W_2(q_1,q_2)=U_2(q_1,q_2)-W_1(q_1)W_1(q_2)$ from
\wdef, $W_2(q_1,q_2)$ takes non-zero values only when $q_1$ and
$q_2$ coincide or next to each other.
Such a cluster property of $W_n$s
can be proven in general as follows.
Consider two sets of positions:
$Q=\{q_1,\cdots,q_m\}, Q'=\{q_{m+1},\cdots, q_{m+m'}\}$.
Suppose these two sets are separated from each other.
We will prove by iteration
that $W_{m+m'}(q_1,\cdots,q_{m+m'})$ is zero
always for such separated $Q$ and $Q'$.
We start with the following equation which is valid because
the two sets are separated:
$$
U_{m+m'}(q_1,\cdots,q_{m+m'})
=U_m(q_1,\cdots,q_m)U_{m'}(q_{m+1},\cdots,q_{m+m'}),
\eqn\uuu
$$
where the definition of the product of $U_m$s follows
that of $W_n$s.
As was shown above, the cluster property is valid for $m=m'=1$.
Assume that $m+m'>2$, and
that $W_n\ (n<m+m')$ satisfies the cluster property.
Then, from the relation \wdef,
$$
\eqalign{
U&_{m+m'}(q_1,\cdots,q_{m+m'})-W_{m+m'}(q_1,\cdots,q_{m+m'})\cr
&=\sum_{k_1\cdots k_{m+m'-1} \atop \sum_{i=1}^{m+m'-1}
ik_i=m+m'}
\left( \prod_{i=1}^{m+m'-1} {1\over k_i!(i!)^{k_i}} \right)
\sum_{\sigma_{m+m'}}
W_1(q_{\sigma_{m+m'}(1)})\cdots
W_{m+m'-1}(\cdots,q_{\sigma_{m+m'}(m+m')})\cr
&=U_m(q_1,\cdots,q_m)U_{m'}(q_{m+1},\cdots,q_{m+m'}),}
\eqn\uwuu
$$
where we used the cluster property of $W_n\ (n<m+m')$
from the second to the last line.
Comparing with \uuu, we obtain $W_{m+m'}(q_1,\cdots,q_{m+m'})=0$.

Comments are in order.
Because of the topological nature of the partition(correlation)
functions of the TLFT,
in the `operator splitting regularization', one has to take care
only of
the topology of the neighborhoods of the operators.
Thus it is rather simple in lower orders, but is
complicated in higher orders since the topology of the neighborhood
becomes complicated.
But we are allowed to choose the neighborhood of an operator
to make calculations as simple as possible.
The difference of the choices will be absorbed in the operators
with orders higher than the considering order.
Consider, for example, a product $W_n(q_1,\cdots,q_n)W_1(q)$,
where the $D$-simplices $q_1,\cdots,q_n$ surround
an area of the lattice.
Naively there are two possibilities of the splitting, say
the inside or the outside of the area.
We should define the inside of the area as a part of
the neighborhood
of $q_1,\cdots,q_n$, because then the splitting is always
done to the outside and this facilitates the calculations.

The partition function $Z(M;\Sigma(c);\delta Z)$ with colored
boundaries $\Sigma(c)$ can be also calculated by the cluster
expansion.
We start with the analogue of \exppar\ and \parope:
$$
\eqalign{
Z(M;\Sigma(c);\delta Z)&=Z(M;U^{\Sigma(c)}),\cr
U^{\Sigma(c)}&=\sum_{n=0}\sum_{q_1,\cdots,q_n}
{1\over n!} U_n^{\Sigma(c)}(q_1,\cdots,q_n),\cr
U_0^{\Sigma(c)}&=\S(c).
}
\eqn\bouexp
$$
Special cares must be taken if some of the $q_1,\cdots,q_n$
are next to $\Sigma$.
Define $W_n^\S$ to single out such boundary effects:
$$
U^{\Sigma(c)}_n(q_1,\cdots,q_n)
=\sum_{m=0}^n
\sum_{\sigma_n} {1\over m!(n-m)!}
W^{\Sigma(c)}_m(q_{\sigma_n(1)},\cdots,q_{\sigma_n(m)})
U_{n-m}(q_{\sigma_n(m+1)},\cdots,q_{\sigma_n(n)}),
\eqn\wsdef
$$
where $U_n(q_1,\cdots,q_n)$ is defined by splitting operators from
$\S$.
One can show that $W^{\Sigma(c)}_n$ satisfies a cluster property,
that is,
$W^{\Sigma(c)}_n(q_1,\cdots,q_n)$ takes zero if
there exists a non-empty subset $Q$ of $\{q_1,\cdots,q_n\}$
separated from its compliment $\bar Q$ as well as
from $\Sigma$ (Fig.\figcluws).
Using the $W^{\Sigma(c)}_n$, the partition function is expressed as
$$
\eqalign{
Z(M;\Sigma(c);\delta Z)&=Z(M;W^{\Sigma(c)}\exp(W)), \cr
W^{\Sigma(c)}&=\sum_{n=0} \sum_{q_1,\cdots,q_n} {1\over n!}
W^{\Sigma(c)}_n(q_1,\cdots,q_n), \cr
W^{\Sigma(c)}_0&=\Sigma(c),
}
\eqn\zms
$$
where the definition of $W$ is given by \wdef\ and \expwpar.

Consider the case that $\Sigma$ is
composed of two boundaries $\Sigma_i\ (i=1,2)$ and
they are separated by $d$ lattices.
Consider a set $Q=\{q_1,\cdots,q_n\}$ for which
$W_n(q_1,\cdots,q_n)\not= 0$.
Then, if $n$ is smaller than the distance $d$,
the set $Q$ is a sum of two sets $Q_1$ and $Q_2$, where
$Q_1$ is in the neighborhood of $\Sigma_1$ but is separated
from $Q_2$ and the boundary $\Sigma_2$, and vice versa for
$Q_2$ (Fig.\figtwobou).
One can prove that, if $n<d$, $W^{\Sigma(c)}_n$ can
be factorized:
$$
\eqalign{
&W^{\Sigma_{12}(c)}_n(q_1,\cdots,q_n)\cr
&=\sum_{m=0}^n \sum_{\sigma_n}
{1\over m!(n-m)!}
W^{\Sigma_1(c_1)}_m(q_{\sigma_n(1)},\cdots,q_{\sigma_n(m)})
W^{\Sigma_2(c_2)}_{n-m}(q_{\sigma_n(m+1)},\cdots,q_{\sigma_n(n)}),
}
\eqn\factor
$$
where $W^{\Sigma_i}_n(q_1,\cdots,q_n)\ (i=1,2)$
has the cluster property that
it is zero if there exists a subset of $\{q_1,\cdots,q_n\}$
separated from the compliment and the boundary $\Sigma_i$.
An important point is that
$W^{\Sigma_i}_n$ is independent of the other boundary.
Substituting \zms\ with \factor, we obtain
$$
Z(M;\Sigma_{12}(c);\delta Z)
=Z(M;(W^{\Sigma_1(c_1)}W^{\Sigma_2(c_2)}+W_d^{\Sigma_{12}(c)})
\exp(W)),
\eqn\wwwd
$$
where $W_d^{\Sigma_{12}(c)}$ is the order of $(\delta Z)^d$.

\noindent
{\it 3.2 First order phase transitions}

In the previous section, we investigated volume-dependent TLFTs,
and showed that a TLFT is on a multiple
first order phase transition point and that the number of the
phases around it is equal to the number of the irreducible TLFTs
of which the TLFT is the direct sum.
This statement is exact because no approximation is used, but
is valid only in the parameter space of a volume-dependent
TLFT.
In this section, we will show, in the approximation of the
first order of the cluster expansion,
the same statement in the parameter space of the general
deformations of the local weights.

The operator $W_1(q)$ is to take away a $D$-simplex and
contract the colorings with those of $\delta Z({\cal T}^D(q);c)$.
Thus the `operator splitting regularization' implies
$$
\eqalign{
Z(M;\Sigma(c),\prod_{i=1}^n W_1(q_i))
&=Z(M;\Sigma(c),B_0(c_1),\cdots,B_0(c_n))
\prod_{i=1}^n g^{c_ic'_i}_{B_0}\delta Z({\cal T}^D(q_i);c'_i)\cr
&=Z(M;\Sigma(c),\phi^{j_1},\cdots,\phi^{j_n})
\prod_{i=1}^n \phi_{j_i}(c_i)g^{c_ic'_i}_{B_0}
\delta Z({\cal T}^D(q_i);c'_i), \cr
}
\eqn\profirclu
$$
where $B_0=\pt{\cal T}^D$ and
$\phi_i$s are the physical states on $S^{D-1}$ .
Here we have used \prophi.
Thus, in the first order of the cluster expansion,
$W$ is
$$
\eqalign{
W&=\sum_q W_1(q)=A\delta Z_i \phi^i, \cr
\delta Z_i&\equiv {1\over A} \sum_q \phi_i(c)
g^{cc'}_{B_0}\delta Z({\cal T}^D(q);c'), \cr
}
\eqn\wval
$$
where $A$ is the total number of the $D$-simplices in $M$,
and $\delta Z_i$s
are the mean values of the deformations of the local weights
projected to the physical states on $S^{D-1}$.

Using \expwpar, \formone, \formtwo\ and \decomp,
$Z(M;\delta Z)$ can be calculated easily
in the canonical basis:
$$
\eqalign{
Z(M;\delta Z)&=Z(M;\exp(W))\cr
&=\sum_{i=1}^N {1\over \lambda^i} Z(M;\phi^i \exp(W))\cr
&=\sum_{i=1}^N{1\over \lambda^i} Z(M;\phi^i
\sum_{n=0} {1\over n!} (A\delta Z_i\phi^i)^n)\cr
&=\sum_{i=1}^N{1\over \lambda^i}
Z^i(M)\exp(A\lambda^i\delta Z_i), \cr
}
\eqn\ziexp
$$
where $N$ is the dimension of the physical Hilbert space on
$S^{D-1}$, or the number of the irreducible TLFTs of which
the TLFT is a direct sum.
This is formally equivalent to \vdtlftpar\ with
$p_i=\lambda^i \delta Z_i$,
and again the TLFT($\delta Z_i=0$) is on a multiple first order
phase transition point with $N$ different phases around it.

In the previous section, the order parameter is given by the
one-sphere expectation value.
Analogously, here we consider the following expectation value:
$$
\langle {1\over \lambda^k} \phi^k \rangle^{M,\delta Z}
\equiv \lim_{A\rightarrow \infty} {1\over \lambda^k}
{Z(M;\delta Z;B_0(c))\phi^k(c) \over Z(M;\delta Z)}.
\eqn\orddef
$$
Using the discussions of the cluster expansion with boundaries in
the previous subsection, we obtain, in the same way as \ziexp,
$$
\eqalign{
Z(M;\delta Z;B_0(c))\phi^k(c)
&=\sum_{j=1}^N Z(M;(\phi^k+{O(\delta Z)^k}_j\phi^j)\exp(W))\cr
&=\sum_{j=1}^N ({\delta^k}_j+{O(\delta Z)^k}_j)Z(M;\phi^j)
\exp(A\lambda^j\delta Z_j),
}
\eqn\modbou
$$
where ${O(\delta Z)^k}_j$ comes from
the correction to the boundary $B_0$ as in \zms.
Thus, substituting \orddef\ with \ziexp\ and \modbou, we obtain
$$
\eqalign{
\langle {1\over \lambda^k} \phi^k \rangle^{M,\delta Z}
&=\lim_{A\rightarrow \infty} {1\over \lambda^k}
{\sum_{j=1}^N({\delta^k}_j+{O(\delta Z)^k}_j)Z(M;\phi^j)
\exp(A\lambda^j\delta Z_j) \over
\sum_{j=1}^N {1\over \lambda^j} Z(M;\phi^j) \exp (A \lambda^j
\delta Z_j)} \cr
&={\delta^k}_m+{O(\delta Z)^k}_m,}
\eqn\evaord
$$
where we assumed that the system is in the $m$-th phase,
that is, $\lambda^m\delta Z_m$ has the greatest real part
among $\lambda^i\delta Z_i\ (i=1,\cdots,N)$.

\noindent
{\it 3.3 Correlation lengths of fluctuations associated to
trivial local operators}

We here discuss the correlation lengths of fluctuations
associated to trivial local operators
in the cluster expansion
and show that they vanish in the limit of TLFT
$\delta Z \rightarrow 0$.
Consider a closed lattice $M$ and make a new lattice $M'$
with two boundaries $\Sigma_{1,2}$
by taking away some of the $D$-simplices in $M$.
We assume that one of the boundaries $\Sigma_1$
is topologically trivial, that is,
$\Sigma_1$ can be surrounded by a $S^{D-1}$ in $M'$
and separated from $\S_2$ by the $S^{D-1}$.
Thus what we consider here
are the fluctuations associated
to such a topologically trivial local boundary $\S_1$.

Let ${\cal O}_{1,2}$
be the operations to make the colored boundaries
$\Sigma_{1,2}(c_{1,2})$, respectively.
As was discussed previously, these operators are modified in
the cluster expansion.
Assume that
each operator is modified to ${\cal O}'_{1,2}$ respectively.
Then the composite operator ${\cal O}_1{\cal O}_2$ is modified to
${\cal O}'_1{\cal O}'_2+{\cal O}_3$,
where the ${\cal O}_3$ is of order $(\delta Z)^d$
when the two boundaries are separated by
$d$ lattices.
Then the expectation value of
the correlation function of the two operators is calculated as
$$
\langle {\cal O}_1 {\cal O}_2 \rangle ^{M,\delta Z}
\equiv
\lim_{A\rightarrow \infty}
{Z(M;{\cal O}_1,{\cal O}_2;\delta Z) \over Z(M;\delta Z)}
=\lim_{A\rightarrow \infty}\left(
{Z(M;{\cal O}'_1{\cal O}'_2\exp(W))\over Z(M;\exp(W))}
+
{Z(M;{\cal O}_3\exp(W))\over Z(M;\exp(W))}
\right),
\eqn\twocor
$$
where $W$ is that in \expwpar.
Since the operator ${\cal O}_1$ is a trivial operator, the first
term can be factorized:
$$
\eqalign{
\lim_{A\rightarrow \infty}
{Z(M;{\cal O}'_1{\cal O}'_2\exp(W))\over Z(M;\exp(W))}
&=
\lim_{A\rightarrow \infty}
{\sum_{k=1}^N
Z(S^D;\phi_k {\cal O}'_1)Z(M;{\cal O}'_2\phi^k\exp(W))
\over
\sum_{k=1}^N {1\over \lambda^k}Z(M;\exp(W)\phi^k)}\cr
&=
\lim_{A\rightarrow \infty}
Z(S^D;\phi_m {\cal O}'_1){Z(M;{\cal O}'_2\phi^m\exp(W))
\over
 {1\over \lambda^m}Z(M;\exp(W)\phi^m)},\cr
}
\eqn\faccor
$$
where we cut $M'$ at the $S^{D-1}$ and used the constraint
\glud, \prophi\ with $\S =S^{D-1}$
and \formtwo\ in the first line, and we assumed that
the system is in the $m$-th phase.
On the other hand, the expectation values of
the one-point function of each operator are
$$
\eqalign{
\langle {\cal O}_1 \rangle^{M,\delta Z}
&=\lim_{A\rightarrow \infty}
{Z(M;\exp(W)\phi^m)Z(S^D;\phi^m{\cal O}'_1)
\over
{1\over \lambda^m} Z(M;\exp(W)\phi^m)}\cr
&=\lim_{A\rightarrow \infty}
\lambda^m Z(S^D;\phi^m{\cal O}'_1) ,\cr
\langle {\cal O}_2 \rangle^{M,\delta Z}
&=\lim_{A\rightarrow \infty}
{Z(M;{\cal O}'_2\exp(W)\phi^m)\over Z(M;\phi^m\exp(W))},\cr
}
\eqn\onecor
$$
where we have used the same cut as above in the first line.
Thus, combining  \twocor, \faccor\ and \onecor,
we obtain the connected correlation function as
$$
\eqalign{
\langle {\cal O}_1 {\cal O}_2 \rangle _c ^{M,\delta Z}
&\equiv
\langle {\cal O}_1 {\cal O}_2 \rangle ^{M,\delta Z}
-
\langle {\cal O}_1 \rangle ^{M,\delta Z}
\langle {\cal O}_2 \rangle ^{M,\delta Z}\cr
&=
\lim_{A\rightarrow \infty}
{Z(M;{\cal O}_3\exp(W))\over Z(M;\exp(W))}
\sim O((\delta Z)^d).
}
\eqn\conooneotwo
$$
This concludes that the correlation length
is of order $-1/\ln (\delta Z)$,
which vanishes in the limit of TLFT $\delta Z\rightarrow 0$.

\noindent
{\it 3.4 An example --- 2D $Q$-state Potts model}

The previous example of 2D lattice QCD cannot be
on a multiple first order phase transition point, if we
respect the physical constraints.
But the 2D Q-state Potts model has a
multiple first order phase transition point of a TLFT in
the physical parameter space\refmark{\BACPET}.

The Potts spins are located on
the 1-simplices, and they interact among nearest neighbors
and with external magnetic fields.
This model is defined by the following local partition function
and gluing tensor:
$$
\eqalign{
g^{ij}=\delta ^{ij},\ C_{ijk}=\kappa W(i,j)W(j,k)W(k,i), \cr
W(i,j)=\exp(\beta(\delta_{ij}-1)+h_i+h_j),
}
\eqn\potdef
$$
where $\beta$ is the inverse temperature, $h_i$s the external
magnetic fields, and the Roman indices run from 1 to $Q$.
As was discussed in the previous example, this model will be a
volume-dependent TLFT if the `flip invariance' is satisfied.
One can easily show that the necessary and sufficient condition of
the invariance is $\beta=0$ or $\infty$\refmark{\BACPET}.

For $\beta=\infty$, $W(i,j)=\delta_{ij} \exp(2h_i)$.
To perform the Laplace decomposition,
we calculate $Z(B^2;A;B_0(i,j,k))$, where $B^2$ denotes a disk
and $B_0(i,j,k)$ a three-gon with coloring $i,j,k$:
$$
Z(B^2;A;B_0(i,j,k))=\delta_{ij}\delta_{ik}
\left(\kappa\exp(6h_i)\right)^A
=\sum_{\alpha=1}^Q \delta^\alpha_i\delta^\alpha_j\delta^\alpha_k
\left(\kappa\exp(6h_\alpha)\right)^A.
\eqn\lapdecofzero
$$
The local weight $\delta^\alpha_i \delta^\alpha_j\delta^\alpha_k$
trivially defines an irreducible TLFT for each $\alpha$.
Thus the point
$\kappa \exp(6h_\alpha)=1$ for all $\alpha$ must be a multiple first
order phase transition point with $Q$ different phases around
it.
On the other hand, since,
after redefining $\kappa$ and the magnetic fields properly,
this point corresponds to the point of zero temperature and
zero magnetic fields of the $Q$-state Potts model, it is a
multiple first order phase transition point
with $Q$ different phases labeled by the directions of the spin.
 Thus the two results coincide.
The cluster expansion of the partition function near the
TLFT has a direct relation
to the low temperature expansion of the Potts model.
This suggests that
the cluster expansion near a TLFT is at best an asymptotic
expansion in general.

For $\beta =0$, $W(i,j)=\exp(h_i+h_j)$.
One obtains easily
$$
Z(B^2;A;B_0(i,j,k))=\kappa(\S_{l=1}^Q\exp(4h_l))^{3/2})^A
(\S_{l=1}^Q\exp(4h_l))^{-3/2}\exp(2(h_i+h_j+h_k)).
\eqn\hightemppart
$$
Here one can see that the local weight
$(\sum_{l=1}^Q\exp(4h_l))^{-3/2}\exp(2(h_i+h_j+h_k))$
is of the form $\varphi_i\varphi_j\varphi_k$ with
$\varphi_i=(\sum_{l=1}^Q\exp(4h_l))^{-1/2}\exp(2h_i)$.
Hence the local weight can be transformed to
the form $\delta^1_i\delta^1_j\delta^1_k$,
and it defines an irreducible TLFT.
This concludes that the TLFT at $\beta=0$ and
$\kappa(\sum_{l=1}^Q\exp(4h_l))^{3/2}=1$
is an irreducible TLFT and is not on a first order phase
transition point.
On the other hand, this point corresponds to the infinite temperature
limit of the Potts model, and hence is not on a first order phase
transition point.
The cluster expansion near the TLFT corresponds to
the high temperature expansion of the Potts model.


\chapter{Neighborhood of three-dimensional Z$_p$ model}

So far we have discussed the physical states on
$S^{D-1}$ of a TLFT and their relations to
the phase structure of first order phase transitions.
We gave some examples in two dimensions.
Since, in two dimensions,
a boundary has always the topology of $S^1$,
all the physical states are those on $S^1$.
Thus only the structure of first order phase transitions
is associated to a two-dimensional TLFT.
But how about the physical states on non-trivial boundaries in
higher dimensions?
We will not give here any general statements about it.
We will discuss it in a simple three-dimensional model on a
manifold with a certain topology.
The method is again the cluster expansion.
But here higher orders will play essential roles.

\noindent
{\it 4.1 Definition and physical states }

The model we discuss here is
a simpler analogue of the model by Turaev and Viro\refmark{\TURVIR}.
We consider the $Z_p$ gauge group in place of
the quantum groups. (See [\DIJWIT] and [\BOU].)

Consider a simplicially decomposed
three-dimensional manifold $M$ possibly with some boundaries.
The degrees of freedom are on its 1-simplices, and they are
integers running from 0 to $p-1$.
The coloring of the lattice is performed by the assignment of
these integers to all the 1-simplices.
The weight for each inner 0-simplex is $1/p$, and that for
each 0-simplex on boundaries is $1/\sqrt{p}$.
The weight for each 3-simplex is given by the $Z_p$ analogue
of $6j$-symbol:
$$
\sym j_1/j_2/j_3/j_4/j_5/j_6/ =
\delta^{(p)}_{j_1+j_2+j_3,0}
\delta^{(p)}_{j_1+j_5-j_6,0}
\delta^{(p)}_{j_2-j_4+j_6,0}
\delta^{(p)}_{j_3+j_4-j_5,0},
\eqn\locweidef
$$
where $\delta^{(p)}_{ij}$ takes 1 for $i=j$ mod $p$, but 0 otherwise.
The weight \locweidef\ is a flatness condition for the gauge group
$Z_p$, that is, the summation of $j_i$s over the 1-simplices
surrounding each 2-simplex should vanish with mod $p$.
The partition function is defined as
$$
Z(M;\S)=p^{-(v_i+v_b/2)} \sum_{j} \prod_{T}
\sym j^T_1/j^T_2/j^T_3/j^T_4/j^T_5/j^T_6/,
\eqn\zppardef
$$
where $v_i$ and $v_b$  are the total
numbers of the inner 0-simplices and
those on the boundaries, respectively.
The summation is over all the colorings, that is,
all the assignments of the $p$ integers to the
1-simplices.
$j^T_i$ represents the six integers associated to each
3-simplex and
the product is taken over all the 3-simplices in $M$.

All the triangulations of the inside of a three-dimensional
manifold can be related by
the two local moves (1,4) and (2,3) and their
inverse moves\refmark{\GROVAR} (Fig.\figmove).
The partition functions \zppardef\ are invariant
under these moves, and
hence they define a TLFT.

The boundary of a 3-manifold is a closed 2-manifold.
Let $B_g$ denote a triangulated surface with genus $g$.
Since \zppardef\ consists of the products of the flatness
conditions, one can show easily that
the basis of the physical Hilbert space on $B_g$
can be labeled by the currents on $\alpha$ and $\beta$
cycles of $B_g$:
${\cal H}_g=\{\{j_{\alpha_i}j_{\beta_i}\ (i=1,\cdots,g) \}
| j_{\alpha_i},j_{\beta_i}=0,\cdots,p-1\}$.
Hence the dimension of the physical Hilbert space is given by
${\rm dim}({\cal H}_g)=p^{2g}$.
Especially, the dimension of the physical Hilbert space
on the sphere is one, so this system is not on a
first order phase transition point.

\noindent
{\it 4.2 Evaluation of the partition function}

We will calculate by the cluster expansion
the partition function of the system obtained by shifting
the local weights by little amounts from those of the
$Z_p$ model.
Since the deformed system is not a TLFT anymore,
we have to fix the lattice.
The lattice we consider is a simplicially
decomposed manifold $M_g$ with the topology of $S^1\times B_g$.
Since we finally consider the
thermo-dynamical limit, we can consider a surface composed of
2-simplices in $M_g$ which cuts $M_g$ almost perpendicularly to
the direction along $S^1$ at any place $t(\in S^1)$. We assume
the followings for the triangulated surface
$t\times B_g$ (Fig.\figgeo):
\nextline
(1) The holes are regularly
ordered and the length of the circumference of each
hole is about $l_1$ lattices. \nextline
(2)The length of the cycle winding
through two neighboring holes is about $l_2$ lattices.\nextline
(3) $l_1 <  l_2$.

In the thermo-dynamical limit, the length of $S^1$ and the extent
of $B_g$ will become so large that
the partition function of the deformed system
can be well approximated by
a correlation function of the original TLFT on $M_g$:
$Z(M_g;\delta Z)=Z(M_g;\exp(W))$.

The first order of the cluster expansion of $W$ is the operator
taking away one 3-simplex. Thus the
`operator splitting regularization' implies
$$
W=Af_0\phi_0,
\eqn\expwsph
$$
where $\phi_0$ is the physical state on $S^{2}$, and $f_0$ is
the order of $\delta Z$.
Let us consider the $p$-th order $(p>1)$.
The $p$-th order of $W$ comes from the $p$ 3-simplices forming
a cluster.
If $p<l_1$, the $p$ 3-simplices can be
always surrounded by a $S^{2}$ in $M_g$.
Taking such a $S^{2}$ of a small size,
one can define the neighborhood of the cluster
as the neighborhood of the ball whose boundary is the $S^{2}$.
Then the `operator splitting regularization' projects the
operation of taking away the $p$ 3-simplices to the
physical state $\phi_0$.
Thus the cluster expansions below the $l_1$-th order
can be included in the form \expwsph.

New operators arise when we consider the $l_1$-th order.
There appear the cases that $l_1$ 3-simplices
form the non-trivial loop surrounding one of the $g$ holes
in $M_g$. Then the 3-simplices
can not be surrounded by a $S^{2}$ in $M_g$.
The simplest definition of the neighborhood of such
3-simplices is to choose a small torus which surrounds
the 3-simplices around the hole.
Then such operations are projected to
the physical states on the torus by the `operator splitting
regularization'.
Thus we obtain
$$
W=Af_0\phi_0+T \sum_{k=1}^g \sum_{i,j=0}^{p-1}f^k_{ij}\phi^{ij}_k,
\eqn\expwtor
$$
where $T$ is the number of the lattices along $S^1$.
The $\phi^{ij}_k$ denotes the physical states on the torus
around the $k$-th hole in $M_g$, and is labeled
by the two integers along the $\alpha$ and $\beta$ cycles of the
torus.
The $f^k_{ij}$s are the mean values of the coefficients
coming from the projections of the operations
to $\phi_k^{ij}$:
$f^k_{ij}={1\over T}\sum_t f^k_{ij}(t)$.
Thus they are the order of $(\delta Z)^{l_1}$.
Since, from \formtwo,
the first term of \expwtor\ is merely the all over
factor of the partition function, we will consider only
the second term.
To calculate $Z(M_g;\exp(W))$, we begin with determining
the `matrix elements' of $\phi^{ij}_k$.
Consider a manifold $T_{B_g}=[0,1]\times B_g$ with an insertion of
the operator $\phi^{ij}_k$.
Then, using similar discussions to obtain the physical Hilbert space
before, the `matrix elements' of $\phi^{ij}_k$
are obtained from the flatness condition as
$$
Z(T_{B_g};\psi_g^{\{i_l,j_l\}*},\phi_k^{ij},\psi_g^{\{i_l',j_l'\}})
=\delta^{(p)}_{i,i_k}\prod_{l=1}^g \delta^{(p)}_{i_l,i_l'}
\delta^{(p)}_{j_l,j_l'+\delta_{kl}j},
\eqn\matphi
$$
where $\psi_g^{\{i_l,j_l\}*}$ and $\psi_g^{\{i_l',j_l'\}}$ are
the physical states on the boundaries $B_g^*$ and $B_g$,
respectively.
Here we took the normalization of the inserted operator
$\phi_k^{ij}$ such that
there appear no numerical factors in the right-hand side of
\matphi.
The case that there are two insertions can be considered by
taking the square of the matrix \matphi, and we obtain
the fusion rule
$$
\phi_k^{ij}\phi_k^{i'j'}\sim \delta^{(p)}_{ii'}\phi_k^{i,j+j'}.
\eqn\fusphi
$$
For simplicity, we will use
$\varphi_k^{ij} \equiv (1/p) \sum_{l=0}^{p-1} \exp(i2\pi jl/p)
\phi_k^{il}$ in place of $\phi_k^{ij}$,
where the $i$ in $i2\pi$ denotes the imaginary unit, and do not
confuse it with the label for the states.
Then the fusion rule and $W$ become
$$
\eqalign{
\varphi_k^{ij}\varphi_k^{i'j'}
&\sim \delta^{(p)}_{ii'}\delta^{(p)}_{jj'} \varphi_k^{ij},\cr
W&=\sum_{k=1}^g \sum_{i,j=0}^{p-1} h_{ij}^k \varphi_k^{ij},
}
\eqn\newope
$$
where $h_{ij}^k$s are the order of $(\delta Z)^{l_1}$.
Then, using the fusion rule \newope\ and
$Z(M_g;\varphi_1^{i_1j_1},\cdots,\varphi_g^{i_gj_g})$ $=1$
obtained by taking the trace of \matphi, we obtain
$$
Z(M_g;\exp(W))
=Z(M_g;\prod_{k=1}^g \sum_{i,j=0}^{p-1} \varphi_k^{ij}
\exp(T h_{ij}^k))
=\prod_{k=1}^g \sum_{i,j=0}^{p-1} \exp(T h_{ij}^k).
\eqn\valpar
$$
Thus the free energy of the system is
$$
F(M_g;\delta Z)
=-\sum_{k=1}^g \ln( \sum_{i,j=0}^{p-1} \exp(T h_{ij}^k)).
\eqn\valfre
$$

\noindent
{\it 4.3 Phase structure}

Since we assume $\delta Z \ll 1$,
if one takes $l_1\rightarrow\infty$ in the thermo-dynamical limit,
the system will become uninteresting.
Thus we fix $l_1$ and $l_2$ in the thermo-dynamical limit.
Since then the number of holes $g$ is proportional to the extent
of $B_g$, it is natural to consider the free energy per
hole and per lattice along $S^1$:
$$
f(\delta Z)
=\lim_{T,g\rightarrow \infty} {1\over gT} F(M_g;\delta Z)
=-{\rm max}_{ij}(h_{ij}),
\eqn\valfre
$$
where ${\rm max}_{ij}(h_{ij})$ is to take the $h_{ij}$ which has
the greatest real part among all the $h_{ij}\ (i,j=0,\cdots,p-1)$,
and we assumed for simplicity that
$h_{ij}^k$ does not depend on the hole:
$h_{ij}^k=h_{ij}$.
{}From \valfre, we conclude that the TLFT is on a $l_1$-th order
discrete phase transition point with $p^2$ different phases
around it.

Comments are in order.
Firstly, the number of the different phases around the TLFT
would depend on the physical constraints of the system.
Secondly, the order of the phase transition $l_1$ is not universal,
since $l_1$ depends heavily on the local lattice structures
around holes.
Thirdly, if $l_1$ is large, the phase transition is very weak
and hence the phase structure might become different because
of possible large fluctuations.
Finally, the order of the phase transitions is $l_1$
only on the point of the TLFT ($\delta Z =0$) in general.
An example is as follows.
Suppose the free energy is given by
$-{\rm max}(x^{q}y^{l_1-q},x^{q'}y^{l_1-q'})\ (q\not = q')$.
Then a phase transition line is at $x=y$.
But the line is a first order phase transition line except
the point $x=y=0$, where the order is $l_1$.

\noindent
{\it 4.4 Correlation length and order parameter}

In the preceding section, we showed that the correlation lengths
of the fluctuations associated to a local trivial operator
vanish in the limit of TLFT. In this section, we investigate
the correlation lengths of the fluctuations associated to
local non-trivial operators in the perturbed $Z_p$ model.
We will show that the correlation lengths of the fluctuations
associated to the operators which wind through two neighboring
holes diverge in the limit of TLFT.

We first discuss the correlation lengths of the fluctuations
associated to the torus around a hole.
Consider a triangulated torus $B_1$ around the
$k$-th hole and an arbitrary
boundary $\Sigma$ obtained by taking away
some 3-simplices from $M_g$. We assume they are separated
by $d$ lattices.
We start with evaluating the following partition function in the
cluster expansion of the $l_1$-th order:
$$
\eqalign{
\lim_{T\rightarrow\infty}\sum_{c'}
Z(M_g;\Sigma(c),B_1(c');\delta Z)\varphi_k^{ij}(c')
&=\lim_{T\rightarrow\infty}\sum_{i',j'=0}^{p-1}
Z(M_g;(\Sigma(c)'{D_k^{ij}}_{i'j'} \varphi_k^{i'j'}
+{\cal O}_d)\exp (W))\cr
&=\lim_{T\rightarrow\infty}
Z(M_g;(\Sigma(c)'{D_k^{ij}}_{mn}\varphi_k^{mn}
+{\cal O}_d)\exp(W))\cr
&=\lim_{T\rightarrow\infty}(
{D_k^{ij}}_{mn}Z(M_g;\Sigma(c);\delta Z)+Z(M_g;{\cal O}_d\exp(W))).
}
\eqn\evalcor
$$
Here the modifications of the operators by the cluster expansion
are considered, and ${D_k^{ij}}_{i'j'}$ are the coefficients
obtained by projecting the modified $B_1(c)$ to the
$\varphi_k^{ij}$s. Hence ${D_k^{ij}}_{i'j'}=\delta^{(p)}_{ii'}
\delta^{(p)}_{jj'}+O(\delta Z)$. We assumed that,
in the limit $T\rightarrow\infty$, $h^k_{mn}$ dominates
among $h^k_{ij}\ (i,j=0,\cdots,p-1)$, and
used $\exp(W)=
\prod_{k=1}^g \sum_{i,j=0}^{p-1} \varphi_k^{ij} \exp(T h_{ij}^k)$
and the fusion rule \newope\ in the last line.
The operator ${\cal O}_d$
is the modification of the composite operator like
$W_d^{\Sigma_{12}(c)}$ in \wwwd,
and is of order $(\delta Z)^d$.
Substituting \evalcor\ with $\Sigma=\emptyset$, we obtain another
relation
$$
\lim_{T\rightarrow\infty}
\sum_{c'} Z(M_g;B_1(c');\delta Z)\varphi_k^{ij}(c')
=\lim_{T\rightarrow\infty}
{D_k^{ij}}_{mn}Z(M_g;\delta Z),
\eqn\evalone
$$
{}From \evalcor\ and \evalone, we obtain
$$
\eqalign{
\langle \S(c) \varphi_k^{ij} \rangle^{M_g,\delta Z}
&\equiv \lim_{T\rightarrow\infty}
{Z(M_g;\Sigma(c),B_1(c');\delta Z)\varphi_k^{ij}(c')
\over Z(M_g;\delta Z)} \cr
&=\langle \S(c) \rangle^{M_g,\delta Z}
\langle \varphi_k^{ij} \rangle^{M_g,\delta Z}
+\langle {\cal O}_d \rangle^{M_g,\delta Z}.\cr
}
\eqn\correl
$$
Hence the correlation lengths of the fluctuations associated
to $B_1$ are the order of $-1/\ln \delta Z$, which vanish
in the limit of the TLFT ($\delta Z \rightarrow 0$).

{}From \evalone, one can see that the order parameters are
given by the following one-$\varphi$ expectation values:
$$
\langle \varphi_k^{ij} \rangle^{M_g,\delta Z}
={D^{ij}_k}_{mn}
=\delta^{(p)}_{im}
\delta^{(p)}_{jn}+O(\delta Z).
\eqn\ordparzep
$$

Next we consider the fluctuations associated to the torus winding
through two neighboring holes. We denote the physical states
associated to the torus by $\Gamma_{kl}^{ij}$,
where $kl$ labels the two neighboring holes and
$ij$ the physical states (Fig.\figopegam).
Using the similar discussions as before,
the `matrix elements' of these operators are
obtained from the flatness condition:
$$
Z(T_{B_g};\psi_g^{\{i_m,j_m\}*},\Gamma_{kl}^{ij},
\psi_g^{\{i_m',j_m'\}})
=\delta^{(p)}_{j,j_k'-j_l'}\prod_{m=1}^g
\delta^{(p)}_{i_m,i_m+\delta_{km}i-\delta_{lm}i}
\delta^{(p)}_{j_m,j_m'},
\eqn\matgam
$$
where we normalized the $\Gamma$s such that there are no numerical
factors in the right-hand side of \matgam.
For convenience, we will use the operators $\gamma_{kl}^{ij} =
\sum_q \Gamma_{kl}^{iq} \exp(i2\pi qj/p)$ in place of $\Gamma$.

One has to take into account the deformations of the operators
in the cluster expansion.
But, since the main contribution comes from the original operators,
we will neglect the deformations. This simplification does not
change the conclusions concerning the correlation lengths.

By using the fusion rule \newope\ and taking the trace of the product
of the matrices \matphi\ and \matgam,
the one-$\gamma$ function is evaluated as
$$
\eqalign{
Z(M_g;\gamma_{kl}^{ij};\delta Z)
=Z(M_g;\gamma_{kl}^{ij}\exp(W))
&=\sum_{\{i_m,j_m\}}
Z(T_{B_g};\psi_g^{\{i_m,j_m\}*},\gamma_{kl}^{ij}\exp(W),
\psi_g^{\{i_m,j_m\}})\cr
&=\delta^{(p)}_{i,0}\delta^{(p)}_{j,0}Z(M_g;\delta Z).
}
\eqn\parofonegam
$$
Thus the one-$\gamma$ expectation value is
$$
\langle \gamma_{kl}^{ij} \rangle^{M_g,\delta Z}
=\delta^{(p)}_{i,0}\delta^{(p)}_{j,0}.
\eqn\expgam
$$

We next consider the two-$\gamma$ function. We insert two
$\gamma$s separated by $t$ lattices in the direction along
$S^1$.
As same as above, the two-$\gamma$ function is
calculated straightforwardly as
$$
\eqalign{
Z&(M_g;\gamma_{kl}^{ij}(t),\gamma_{k'l'}^{i'j'}(0);\delta Z)\cr
&=\sum_{\{i_m,j_m\}\{i_m',j_m'\}}
Z(T_{B_g};\psi_g^{\{i_m,j_m\}*},\exp(W(T-t))\gamma_{kl}^{ij},
\psi_g^{\{i_m',j_m'\}})\cr
&\ \ \ \ \ \ \ \ \ \ \ \ \ \ \ \ \ \ \ \ \ \ \ \ \ \ \ \ \ \ \ \times
Z(T_{B_g};\psi_g^{\{i_m',j_m'\}*},\exp(W(t))\gamma_{k'l'}^{i'j'},
\psi_g^{\{i_m,j_m\}})\cr
&=p^{-2g}\sum_{\{i_m,j_m\}\{i_m',j_m'\}}
\exp(i2\pi(j'(j_{k'}'-j_{l'}')+j(j_k'-j_l'))/p))\cr
&\times \prod_{m=1}^g
\delta^{(p)}_{{i_m}',i_m+\delta^{(p)}_{mk'}i'-\delta^{(p)}_{ml'}i'}
\delta^{(p)}_{{i_m},{i_m}'+\delta^{(p)}_{mk}i-\delta^{(p)}_{ml}i}
H(t;m;{i_m}',{j_m}'-j_m)H(T-t;m;{i_m},j_m-{j_m}'),
}
\eqn\twogam
$$
where $W(T-t)$ and $W(t)$ are the short expressions for
the $W$s for the intervals $[t,T]$ and $[0,t]$, respectively.
We assumed for simplicity that the $h^k_{ij}$s at $[0,t]$ and
$[t,T]$ are equal, and the $H$ is  defined as
$H(t;m;i,j)\equiv \sum_{k=0}^{p-1} \exp(th^m_{ik}+i2\pi kj/p)$.

To evaluate \twogam\ further, we will consider two cases separately.
First consider the case $k=k',\ l=l'$.
Then the Kronecker's deltas in \twogam\ requires $i=-i',\ j=-j'$,
and we obtain
$$
\eqalign{
Z&(M_g;\gamma_{kl}^{ij}(t),\gamma_{kl}^{i'j'}(0);\delta Z)\cr
&=Z(M_g;\delta Z)
\delta^{(p)}_{i+i',0}\delta^{(p)}_{j+j',0}
(\sum_{r,s=0}^{p-1} \exp(Th^k_{rs}))^{-1}
(\sum_{r,s=0}^{p-1} \exp(Th^l_{rs}))^{-1}\cr
&\times
(\sum_{r,s=0}^{p-1} \exp(th^k_{r-i,s}+(T-t)h^k_{r,s-j}))
(\sum_{r,s=0}^{p-1} \exp(th^l_{r+i,s}+(T-t)h^l_{r,s+j})).\cr
}
\eqn\evatg
$$
In the thermo-dynamical limit $T\rightarrow \infty$,
we fix the separation $t$.
Suppose $h^k_{m_k,n_k}$ and $h^l_{m_l,n_l}$
have the largest real values among $h^k_{r,s}$s and $h^l_{r,s}$s
$(r,s=0,\cdots,p-1)$, respectively.
Then we obtain in the thermo-dynamical limit
$$
\langle \gamma_{kl}^{ij}(t)\gamma_{kl}^{i'j'}(0) \rangle
=\delta^{(p)}_{i+i',0}\delta^{(p)}_{j+j',0}
\exp(-t(h^k_{m_k,n_k}+h^l_{m_l,n_l}-h^k_{m_k-i,n_k+j}
-h^l_{m_l+i,n_l-j})).
\eqn\evatge
$$
Combining \evatge\ and \expgam, one can see
that the correlation lengths associated to the $\gamma_{kl}^{ij}$
($(i,j)\not=(0,0)$) are of order $(\delta Z)^{-l_1}$.
Then, in the limit of TLFT, these correlation lengths diverge.

Next consider the case $(k,l)\not= (k',l')$.
Since then the Kronecker's
deltas in \twogam\ restrict $i$ and $i'$ to be zero, we obtain
$$
Z(M_g;\gamma_{kl}^{ij}(t),\gamma_{k'l'}^{i'j'}(0);\delta Z)
=\delta^{(p)}_{i,0}\delta^{(p)}_{j,0}
\delta^{(p)}_{i',0}\delta^{(p)}_{j',0} Z(M_g;\delta Z).
\eqn\evatwogam
$$
Hence,
$$
\langle \gamma_{kl}^{ij}\gamma_{k'l'}^{i'j'}\rangle^{M_g,\delta Z}
=
\langle \gamma_{kl}^{ij}\rangle^{M_g,\delta Z}
\langle \gamma_{k'l'}^{i'j'}\rangle^{M_g,\delta Z}.
\eqn\evatwogamexp
$$
This implies that
the non-zero contributions to the connected
two $\gamma$ functions of $(k,l)\neq (k',l')$
will come only from the cluster deformations
of the composite operators, and hence are of order $(\delta Z)^d$ if
the two $\gamma$s are separated by $d$ lattices in the direction
perpendicular to the $S^1$.
Thus the correlation lengths to that direction are of order
$-1/\ln \delta Z$, which vanish in the limit of the TLFT.

We have obtained that, in the limit of the TLFT,
the fluctuations associated to the operator $\gamma$ have
the infinite correlation lengths in the direction along $S^1$,
but have zero correlation lengths in the directions
perpendicular to $S^1$.
The fluctuations propagate only in the one-dimensional direction
along the holes.
We could not find any local fluctuations which propagate
in all directions in $M_g$ in the perturbed $Z_p$ model.


\chapter{Summary, comments and discussions}

We investigated the thermo-dynamical natures near TLFTs,
and found that they are in general
on discrete phase transition points, and that
they are on fixed points of
renormalization group transformations at least in a restricted sense.

First we discussed the decomposition of a
volume-dependent TLFT satisfying a certain condition
and showed that it is a direct sum of
irreducible TLFTs with volume-dependent numerical factors.
In the parameter space of the volume-dependent TLFTs we can
discuss the phase structure and the renormalization group flow
exactly, and showed that a TLFT is on a fixed point of
the flow and in general on a multiple first order phase transition
point.
The number of the different phases around a TLFT is equal to
the number of the irreducible TLFTs of which the TLFT is
a direct sum.

To generalize the discussion out of the parameter space of the
volume-dependent TLFTs, we introduced a kind of cluster expansion.
By this perturbative scheme we found the same phase structure
around a TLFT in the first order of the cluster expansion.
Higher orders do not change the result qualitatively
provided the base manifold is trivial.

To specify the roles of the non-trivial topology of the base manifold
and the physical states on non-trivial $(D-1)$-dimensional
boundaries,
we investigated the neighborhood of the $Z_p$ analogue of
Turaev-Viro model in three dimensions.
Then we found another phase structure, that is, the TLFT is on a
higher order discrete phase transition point controlled
by the physical states on boundaries with non-trivial topologies.

We also studied the correlation lengths of fluctuations
associated to various boundaries in the cluster expansion
near TLFTs.
The correlation lengths
of the fluctuations associated to trivial local operators
were shown to be zero in the limit of TLFT.
On the other hand, in the $Z_p$ model, we found the fluctuations
whose correlation lengths diverge in the limit of TLFT.
But these fluctuations propagate only along the topological defects.
In this sense, a TLFT is on a point where
only the fluctuations propagating along
topological defects remain and have infinite correlation lengths.

Comments are in order. We restricted our discussions in the cases
that the numbers of the elements of the index sets $X_l$
are finite.
This restriction makes the numerical factors of the three-sphere
functions $\lambda_i$ to be non-zero, and then we could obtain the
decomposition theorem of a TLFT.
But some continuum formulations of TFTs might allow $\lambda$
to be zero. The twisted $N=2$ Landau-Ginzburg model\refmark{\LGmodel}
with a certain super-potential is such a case.
The TLFT corresponding to the model
is discussed in the paper
by Fukuma, Hosono and Kawai\refmark{\FUKKAW},
where a certain infinite limit must be taken to show the
correspondence.
Moreover it was pointed out\refmark{\BOU,\CHUFUK}
that the zero-coupling limit
of 3D lattice QCD corresponds to the Ponzano-Regge model.
Can we apply our arguments around the Ponzano-Regge model?
It would be too stupid to argue that there are no local
long-range fluctuations just around the zero-coupling limit
of the 3D lattice QCD.
Thus we insist that our discussions are applicable only to
TLFTs with finite degrees of freedom, and
new discussions are needed for the other types of TLFTs and TFTs.

As we have shown,
the physical states on $S^{D-1}$ of a TLFT
label the phases around a multiple
first order phase transition point.
The continuum correspondences of the physical states on $S^{D-1}$
would be scalar observables of TFTs. Thus one could expect
that such scalar observables would have special roles in the
dynamics of TFTs such as `spontaneous symmetry breaking'.
Such discussions were done
by Kawamoto and Watabiki\refmark{\KAWWAT} in their
generalized Chern-Simons theory, which contains fields of forms
of all orders.
Here we cannot comment anything about it,
since the applicability
of our results is obscure as we mentioned in the last paragraph.
Relating TLFTs to TFTs and developing analogous discussions
for TFTs remain as future problems.

The thermo-dynamical natures near a fixed point of a
renormalization group transformation on a continuous phase transition
surface can be well analyzed by the renormalization group technique
and renormalizable field theories.
This is because of the universality, or because some finite number
of interacting long-range modes with some finite number of couplings
control the system essentially.
How about discrete phase transitions?
In general we cannot expect universalities around such phase
transition surfaces, so predictions are difficult.
But when we analyzed the thermo-dynamical natures near TLFTs by
the cluster expansion, the deviations could be projected to
the physical states of TLFTs, and we could obtain some
qualitative natures near TLFTs.
In this sense, the thermo-dynamical natures near TLFTs are
described by the physical observables of TLFTs.
If there were a map to transform any point near a discrete phase
transition surface to the neighborhood of a TLFT,
one could talk about general discrete phase transitions.
But the existence of such a map is obscure. Some difficulties of
the renormalization group technique near a first order
phase transition surface were pointed out\refmark{\EFS}.
We can define a renormalization group flow in the parameter space
of the volume-dependent TLFTs, but will have to overcome
difficulties in the outside.
TLFTs can generate various systems with discrete phase transitions,
but whether they are useful in the investigations of discrete
phase transitions is not clear.

\ack
We would like to thank H.~Kawai, N.~Kawamoto and A.~Yamada
for encouragement and
the members of the elementary particle
sections of Tokyo University for their hospitality.
And we would like to thank H.~Kawai and N.~Kawamoto
for reading this manuscript.
The author is supported by the fellowship
of the Japanese Society for the Promotion of Science
for Japanese Junior Scientists, and this work is
supported by the Grant-in-Aid for Scientific
Research from the Ministry of Education No.~04-1324.

\refout
\endpage
\figout

\bye